This is the authors‘ version (post peer-review) of the manuscript:

Y. Sozen et al. ACS Nano 2026, 20 (24), 17584–17596

# Scalable conformal electronics based on roll-to-roll exfoliated van der Waals semiconductors

*Yigit Sozen[1*], Esteban Zamora-Amo[1], Juan J. Riquelme[1], Andres Castellanos-Gomez[1*]*

*[1]2D Foundry Research Group. Instituto de Ciencia de Materiales de Madrid (ICMM-CSIC), Madrid, E-28049, Spain.*

*corresponding authors yigit.sozen@csic.es, andres.castellanos@csic.es

ABSTRACT

Integrating electronic devices onto surfaces with complex topography such as skin, textiles, and biological tissues requires fabrication strategies that combine mechanical conformability with high electronic performance and scalable manufacturing. While two-dimensional (2D) semiconductors are promising materials for such applications, their integration into conformal electronic systems remains challenging because scalable liquid-phase processing typically yields films with limited electronic performance, whereas high-quality CVD materials require complex synthesis and transfer processes. Here we establish a scalable route toward conformal electronics based on semiconducting van der Waals materials by combining high-throughput roll-to-roll mechanical exfoliation with commercially available temporary tattoo and waterslide decal transfer substrates. This approach enables the fabrication of ultrathin $MoS_2$-based electronic devices that can be transferred onto rough and curved surfaces such as skin, synthetic leather, and plant leaves. The resulting devices operate reliably after transfer and exhibit strong electronic and optoelectronic performance, including photodetectors with responsivities up to ~3.5 A $W^{-1}$, thermistors with temperature coefficients of resistance of −2 to –3.5 % °$C^{-1}$, and ionic-gel-gated field-effect transistors with mobilities reaching ~18 $cm^2$ $V^{-1}$ $s^{-1}$.

## INTRODUCTION

The rapid development of flexible and conformal electronics has enabled the integration of electronic devices onto surfaces with complex topography, including human skin, textiles, and biological tissues[1–4]. Such systems are central to emerging technologies in wearable health monitoring, human–machine interfaces, and bio-integrated sensing, where devices must intimately adapt to soft, curved, and dynamically deforming surfaces while maintaining reliable electrical performance.

Two-dimensional (2D) van der Waals materials are particularly attractive for these applications because their atomic-scale thickness, mechanical flexibility, and absence of dangling bonds enable exceptional mechanical compliance and conformal contact with irregular surfaces.[5–9] These properties allow 2D materials to preserve their electronic and optoelectronic functionality even under extreme bending or when interfaced with soft biological substrates. To date, most demonstrations of conformal bioelectronics based on 2D materials have relied on graphene, which has enabled highly sensitive epidermal sensors and bioelectronic interfaces.[6,10–12] However, the absence of an intrinsic band gap in graphene limits its use in active electronic and optoelectronic components that require efficient switching or strong photoresponse. In contrast, semiconducting transition metal dichalcogenides (TMDs), such as $MoS_2$, possess intrinsic band gaps together with strong light-matter interaction and good carrier mobility, making them attractive materials for active devices including transistors, photodetectors, and chemical sensors.[13–17] Despite these advantages, the integration of semiconducting 2D materials into conformal and skin-interfaced electronic systems remains limited.

A major obstacle lies in the lack of scalable fabrication strategies that simultaneously provide high electronic quality, low cost, and large-area processability. Liquid-phase exfoliation enables scalable production of 2D materials and has been widely used to fabricate printed electronic devices, but the resulting films often exhibit limited electronic performance due to poor interflake connectivity and residual solvents trapped between nanosheets.[18–20] At the other extreme, chemical vapour deposition (CVD) can produce high-quality continuous films with excellent electronic properties, yet it requires sophisticated infrastructure and typically involves complex transfer processes that complicate integration onto unconventional substrates.[21,22] Consequently, most demonstrations of conformal electronics based on semiconducting 2D materials remain limited to small-scale laboratory fabrication approaches.

Temporary tattoo transfer papers have recently emerged as attractive substrates for conformal electronics because they enable the transfer of ultrathin polymer-supported films onto curved and rough surfaces such as skin, glass, or plastics. These systems typically consist of a thin transferable polymer layer supported by a water-soluble sacrificial layer that releases upon wetting, allowing the film to conformally adhere to the target surface. Previous studies have successfully combined tattoo transfer substrates with organic semiconductors to fabricate epidermal electronic devices for physiological monitoring, as well as transferable components such as photodiodes, field-effect transistors, and solar cells.[23–28] While these demonstrations highlight the versatility of tattoo-based transfer strategies for conformal electronics, they have largely relied on organic electronic materials, whose relatively poor charge transport properties limit device performance.

Here we address these limitations by combining high-throughput roll-to-roll mechanical exfoliation of van der Waals materials with ultrathin temporary tattoo and waterslide decal transfer substrates to establish a scalable route toward conformal electronics based on semiconducting 2D materials. Using a recently reported roll-to-roll–like dry exfoliation strategy,[29] this approach produces large-area films composed of interconnected 2D semiconductor flakes with electronic properties superior to those typically obtained from solution-processed materials. These films are integrated into ultrathin transferable platforms using commercially available decal-based substrates, enabling the fabrication of conformal photodetectors, thermistors, and ionic-gel-gated field-effect transistors that can be transferred directly onto rough and curved surfaces such as skin, synthetic leather, and plant leaves. The resulting devices exhibit high responsivity (~3.5 A $W^{-1}$), a large temperature coefficient of resistance (TCR) (from –2 to –3.5 % $°C^{-1}$) within the physiological range, and low-voltage transistor operation with mobilities reaching up to ~18 $cm^2$ $V^{-1}$ $s^{-1}$. By combining scalable production of semiconducting van der Waals materials with simple ultraconformal transfer strategies, this work extends tattoo-based electronics beyond organic systems and establishes a practical platform for high-performance wearable and bio-interfaced devices.

## RESULTS AND DISCUSSIONS

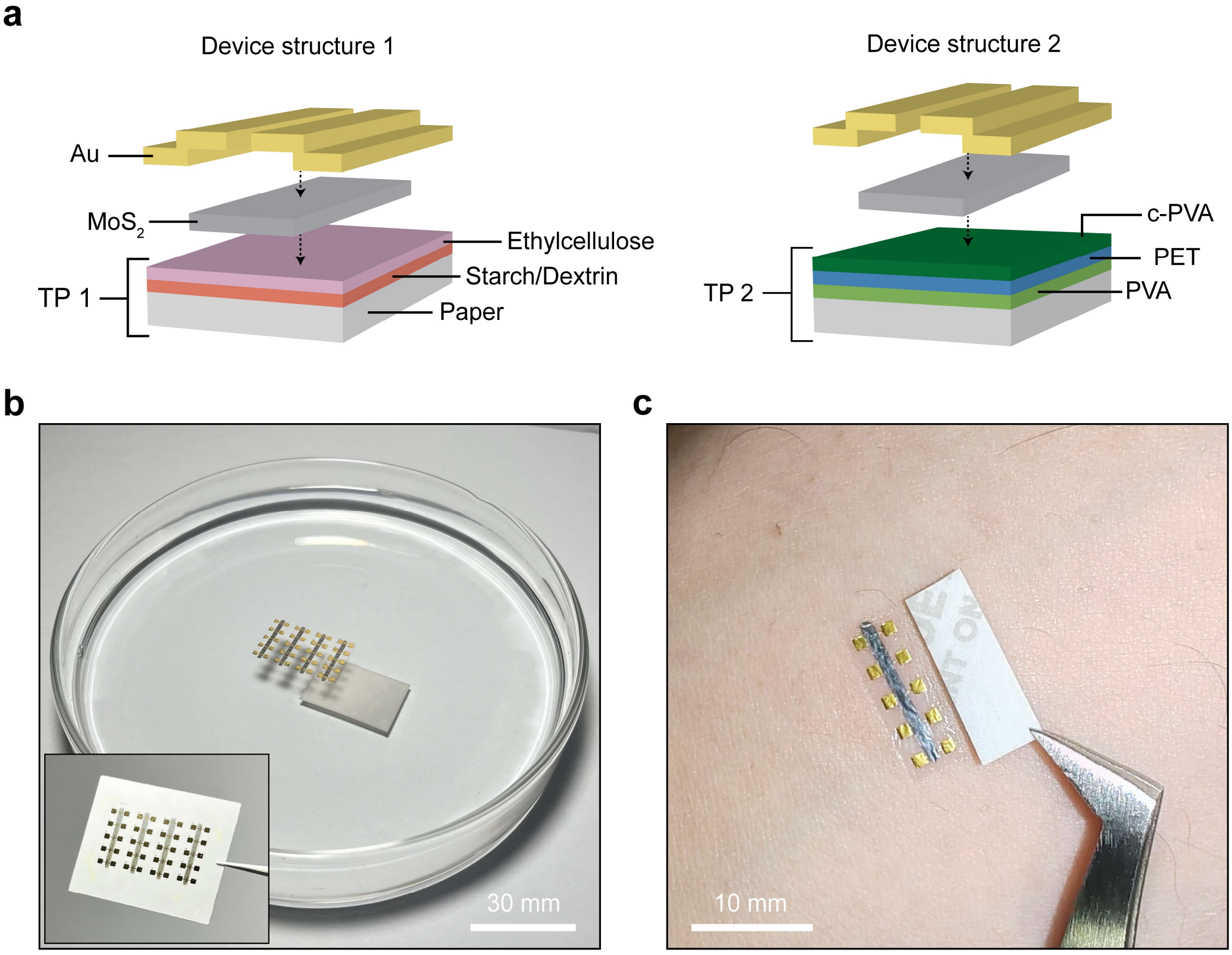

**Figure 1. Delamination and transfer process for $MoS_2$ devices fabricated on tattoo and decal papers.** (a) Cross-sectional schematic of the two different device structures composed of two different transfer papers: TheMagicTouch Tattoo 2.1 paper (left) and Hayes waterslide decal paper (right). In TheMagicTouch Tattoo 2.1 transfer paper, the device sits on a thin ethylcellulose film that can be released from the backing paper through a water-soluble starch/dextrin layer. In Hayes transfer paper, the device is most likely (the detailed proprietary composition is not disclosed by the manufacturer) supported by a partially cross-linked PVA layer (typically designed to increase ink adhesion) on a PET hydrophobic carrier substrate that can be released from the backing paper through a water-soluble PVA film. (b) Delamination of the fabricated $MoS_2$-based devices upon immersion in water using TheMagicTouch Tattoo 2.1 paper. The inset shows the as-fabricated devices on the paper before delamination. The floating device can then be scooped and transferred onto a substrate. (c) Transfer process for Hayes waterslide decal paper. The device is placed on the skin, moistened with a damp cloth, and the water-soaked backing paper is gently slid away to leave the decal adhered to the surface. More details about the transfer process are shown in Figure 2.

The fabrication of ultra-conformal electronic devices was realized through the integration of mechanically exfoliated $MoS_2$ thin films with commercially available transfer papers. Specifically, two complementary transfer media were used: temporary tattoo transfer paper (TheMagicTouch 2.1 Tattoo) and waterslide decal paper (Hayes). Both papers are designed to release a thin hydrophobic polymer film upon wetting, enabling the transfer of printed patterns onto curved and rough surfaces. Notably, in this work the electronic devices are fabricated on the transferable film itself and then transferred onto the target surface, rather than transferring a printed pattern.

Tattoo paper and waterslide decal paper rely on a similar design that involves a hydrophilic backing paper providing mechanical support, a water-soluble sacrificial layer enabling release, and a thin transferable hydrophobic polymer film carrying the printed or functional structures. Despite this similar architecture, the two transfer media differ in composition. The structure and layer composition of tattoo paper and waterslide decal paper used in this study were represented in Figure 1a. From top to bottom, the tattoo paper (TheMagicTouch 2.1 Tattoo) consists of a thin ethylcellulose film, a water-soluble starch/dextrin layer, and a hydrophilic backing paper. Upon soaking, the intermediate layer dissolves and releases the ethylcellulose film. Waterslide decal paper (Hayes) also consists of a multilayer structure, although the manufacturer does not disclose a detailed layer-by-layer description. However, previous patents on inkjet waterslide media suggest a generic architecture comprising a backing paper, a water-soluble poly(vinyl alcohol) (PVA)-based release layer, and a hydrophobic carrier film[30,31]. According to our Raman measurements (see Figure S1), the spectra of the released Hayes film show the characteristic peaks of PVA and polyethylene terephthalate (PET). We therefore infer that Hayes paper follows a similar design: a backing paper, a PVA water-soluble sacrificial layer and a PET hydrophobic carrier film and an ink-receptive layer, most likely composed of cross-linked PVA.

The surface morphology and thickness of transferable films were analysed via atomic force microscopy (AFM) (see Figure S2). For ethylcellulose layer, we obtained an average thickness of 600 nm, which is consistent with the previously reported value[27]. In contrast, the transferred cross-linked PVA/PET film obtained from the Hayes paper is thicker and displays local thickness variations ranging from 1.8 to 2.2 μm. This inhomogeneity is inherent to the waterslide transfer process and is attributed to residual water-soluble PVA remaining on the transferred film. During transfer, the decal is applied face-down, such that the cross-linked PVA-based layer contacts the substrate while the water-soluble PVA layer remains on top after release. Although this upper PVA layer partially dissolves during soaking, it is not completely removed upon detachment of the backing paper, leading to spatial variations in the final film thickness. A detailed description of the waterslide mechanism is provided below (see the discussion related to Figure 2).

The top polymer layer in both tattoo paper and waterslide decal paper served as a carrier platform for device fabrication, enabling subsequent transfer onto target surfaces. Device fabrication begins first with the deposition of $MoS_2$ thin films on the transfer media, which serve as a sensing (active) layer in our devices. These films were prepared via roll-to-roll mechanical exfoliation, a semi-automated process that enables continuous exfoliation of van der Waals crystals to produce large-area films from nanosheets.[29,32] A well-connected percolating network of $MoS_2$ nanosheets was achieved by performing successive film transfers on substrates by a thermal release process. Further details of the exfoliation procedure and transfer process are provided in the "Materials and Methods" section. Optical microscope images obtained after each transfer step are shown in Figure S3, demonstrating the progressive formation of the flake network on each substrate during successive transfers. We performed a quantitative analysis of the evolution of the $MoS_2$ films on each transfer medium by extracting the coverage ratio after each transfer step. As shown in the coverage ratio plots in Figure S4, tattoo paper reaches near-complete surface coverage with fewer transfer steps than Hayes waterslide decal paper. Accordingly, 5–6 and 7–8 transfer steps were required to obtain films with long-range percolating flake networks on tattoo paper and waterslide decal paper, respectively.

AFM topography images of $MoS_2$ nanosheets after a single transfer onto tattoo and waterslide decal paper are presented in Figures S5a and S5b, respectively. The flake lengths were extracted from these images, and the corresponding distributions are shown in Figures S5c and S5d. Fitting these distributions to a lognormal function yields modal nanosheet lengths of 0.8 μm and 0.7 μm for tattoo paper and waterslide decal paper, respectively, with a mean value of 1.3 μm for both. These values are consistent with those previously reported for roll-to-roll exfoliated van der Waals crystals transferred onto rigid $Si/SiO_2$ substrates.[29,32] The similar flake size distributions obtained on substrates with different surface properties (e.g., roughness and elasticity) suggest that the dimensions of the nanosheets are determined mainly by the exfoliation dynamics rather than the substrate characteristics. Due to the surface roughness of the underlying substrates, AFM measurements did not allow reliable determination of individual flake thicknesses. Nevertheless, comparable thickness values to those previously obtained on $SiO_2/Si$ substrates can still be expected, with reported modal thicknesses in the range of approximately 30-40 nm.[29,32] Figure S6 shows Raman spectra acquired from $MoS_2$ flakes transferred onto tattoo and waterslide decal papers, which exhibit two prominent peaks corresponding to $E_{2g}^{1}$ and $A_{1g}$ modes, located at 382 $cm^{-1}$ and 407 $cm^{-1}$, respectively.[33,34] The absence of any noticeable peak shift or broadening indicates that the structural integrity of the $MoS_2$ flakes is well preserved after transfer.

Transfer length method measurements were conducted on $MoS_2$ films with varying numbers of transfer cycles on tattoo paper to evaluate the effect of successive transfers on their electrical conductivity. For this purpose, films consisting of 2, 3, 4, and 5 transfer cycles with a width ($W$) of ~1 mm were prepared on tattoo paper, followed by the deposition of Au contact arrays with varying inter-electrode spacings via thermal evaporation. Figure S7a gives resistance versus channel length plots for films with different numbers of transfer cycles. In each dataset, the resistance decreases with decreasing channel length, while an overall reduction in resistance is observed with an increasing number of transfers, reflecting improved electrical conductivity arising from the progressive formation of percolating pathways across the film. The sheet resistance ($R_s$) for each dataset was extracted from the slope of linear fits, where the slope corresponds to $R_s/W$. As shown in Figure S7b, the sheet resistance exhibits a decreasing trend with an increasing number of transfer cycles. After five transfers, the sheet resistance reaches $\sim 10^9\ \Omega\ \square^{-1}$, which is in good agreement with values reported for $MoS_2$ networks fabricated using other cost-effective techniques, such as liquid-phase exfoliation and abrasion-induced deposition.[35–37]

The final device structures were obtained by transferring bar-shaped $MoS_2$ films onto the transfer papers, followed by the deposition of 80 nm thick Au contacts via thermal evaporation using a commercial shadow mask (Ossila). The final configuration of the fabricated device arrays is shown in the inset of Figure 1b. For the transfer of the fabricated devices onto target surfaces, we followed two distinct transfer processes for tattoo paper and waterslide decal paper. The transferable layer in tattoo paper was detached from the entire structure by soaking it in a water (see Figure 1b) and then transferred onto the target surface by scooping. For waterslide transfer, the decal paper was placed face-down on the skin and wetted using a damp sponge or tissue. The water dissolves the underlying release layer, allowing the paper backing to slide off and leaving the polymer film adhered to the skin (Figure 1c).

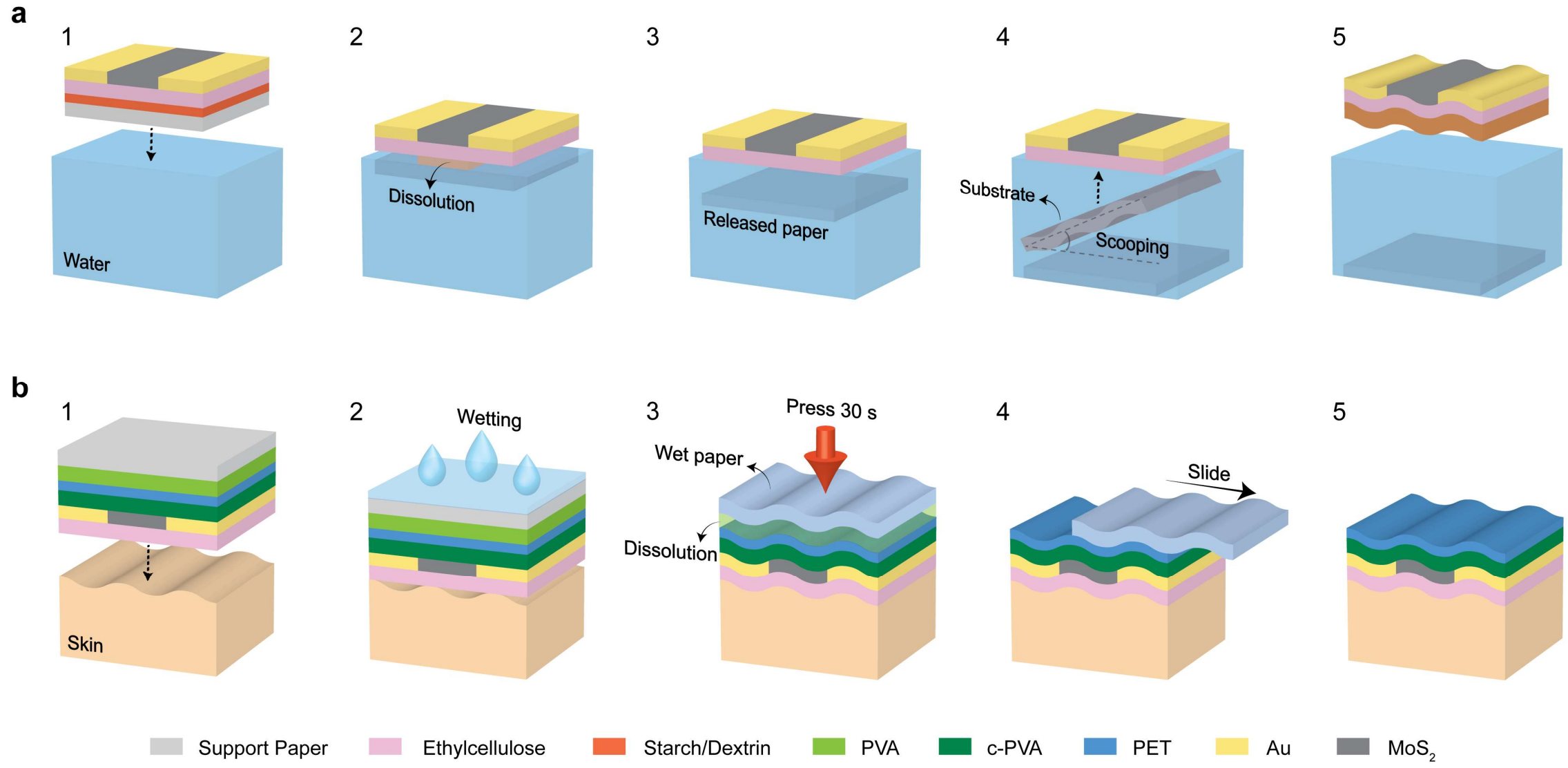

**Figure 2: Step-by-step transfer methods for $MoS_2$-based tattoo devices using TheMagicTouch Tattoo 2.1 paper and Hayes waterslide decal paper.** (a) The transfer process using TheMagicTouch Tattoo 2.1 paper involves immersing the paper in water (1), allowing the fabricated device to delaminate and float (2-3). The floating device is then carefully scooped at a tilted angle and placed onto the target substrate (4-5). (b) The transfer process for Hayes waterslide decal paper starts with adhering the ethylcellulose-encapsulated device to the target substrate (1-2). After wetting, the backing paper is pressed for 30 seconds (3), allowing the PVA layer to partially dissolve (4), and then gently slid off, leaving the device on the target surface (5).

Figure 2a and 2b show step-by-step schematic illustrations of the transfer processes used for the integration of devices fabricated on the respective papers onto target surfaces. We further address the reader to Video S1 and S2 (Supporting Information) for demonstrations of the transfer procedures. In the case of tattoo paper (Figure 2a), the fabricated sample was placed in a Petri dish filled with deionized water and left to soak for approximately 1–2 minutes. During this period, the water-soluble sacrificial layer (starch/dextrin) gradually dissolves, allowing the backing paper to detach and sink. As a result, the freestanding, transferable ethylcellulose film carrying the device starts to float on the water surface and becomes ready for transfer onto the target substrate. After the release, the desired target substrate was carefully immersed beneath the floating film using tweezers and aligned with respect to its position. The device film was then transferred onto the substrate by slowly lifting it through the water surface at a tilted angle of approximately 60°, allowing the film to smoothly adhere without trapping air bubbles or causing mechanical deformation. To further enhance conformal contact of the film with the substrate, a gentle flow of nitrogen gas was applied, ensuring adhesion and minimizing wrinkles or trapped water at the film/substrate interface. Finally, the sample was annealed at 70 °C for 10 min under ambient conditions to remove residual water and ensure full conformal contact with the transferred surface.

Prior to transferring devices from the waterslide decal paper, the device-containing layer was encapsulated with an ethylcellulose layer obtained from tattoo paper (TheMagicTouch 2.1). This additional layer improves the adhesion of the transferred structure to the target substrate, particularly on rough surfaces. The encapsulation step can be carried out using the same face-down transfer procedure described below. Alternatively, devices can be directly transferred without the ethylcellulose overlayer, still yielding fully functional devices, although with reduced adhesion.

The transfer of devices from waterslide decal paper was achieved by placing the transferable, ethylcellulose-covered device-containing side face-down onto the target substrate (see Figure 2b). After positioning, the backing paper was wetted with a small amount of deionized water (typically via a damp sponge or tissue) and gently pressed onto the target surface. After ~30 seconds, the sacrificial layer (PVA) partially dissolves, allowing the backing paper to be gently slid away, while the transfer layer remains securely adhered to the target surface. The same nitrogen-assisted drying and annealing procedure can also be applied to samples transferred onto substrates other than skin. It should be noted that, since the devices are sandwiched between the transferable layer and the target substrate after the transfer, direct electrical access to the electrodes is obstructed. Therefore, prior to the transfer, electrode regions should be pierced with a fine needle to obtain small vias. After transferring the device onto the target substrate, silver paste was applied to these openings to establish electrical contact between the device electrodes and the external contacts.

It should be noted that the scooping method used for tattoo paper is not applicable to waterslide decal paper, as complete self-delamination of the backing layer does not occur upon exposure to water. Conversely, applying the face-down transfer method to tattoo paper may induce mechanical damage, such as cracking of the gold electrodes, due to the ultrathin nature of the ethylcellulose layer. Therefore, each transfer method is specifically suited to its corresponding transfer paper and enables high transfer efficiency.

The two transfer approaches serve distinct and complementary roles, each offering specific advantages and limitations. The scooping transfer process used for devices fabricated on tattoo paper enables the gentle release of the device-containing ethylcellulose film with minimal mechanical stress, making it particularly suitable for fragile and ultrathin device structures. When the film is carefully lifted with the target substrate, wrinkle formation and air trapping at the film–substrate interface can be minimized. However, this method requires careful handling, as the floating film may drift on the water surface during the scooping step, requiring repeated repositioning to achieve precise alignment. In contrast, the waterslide decal paper transfer method is simpler and more direct to implement. It enables straightforward positioning on the target substrate, making it well suited for rapid and routine transfer processes, particularly in applications requiring simple handling and ease of use.

Accordingly, the selection of an appropriate transfer substrate can be critical for specific applications to achieve reliable device integration and optimal performance. For instance, devices fabricated on tattoo paper are thinner than those fabricated on waterslide decal paper, resulting in improved conformal adhesion and reduced risk of delamination. This makes them more suitable for applications requiring intimate mechanical coupling, such as skin-mounted or flexible surface-integrated devices. In contrast, devices transferred using waterslide decal paper without an ethylcellulose overlayer allow direct contact between the device channel and the target surface. This feature is particularly advantageous for wearable biosensing applications, as it enables real-time monitoring of biochemical markers.

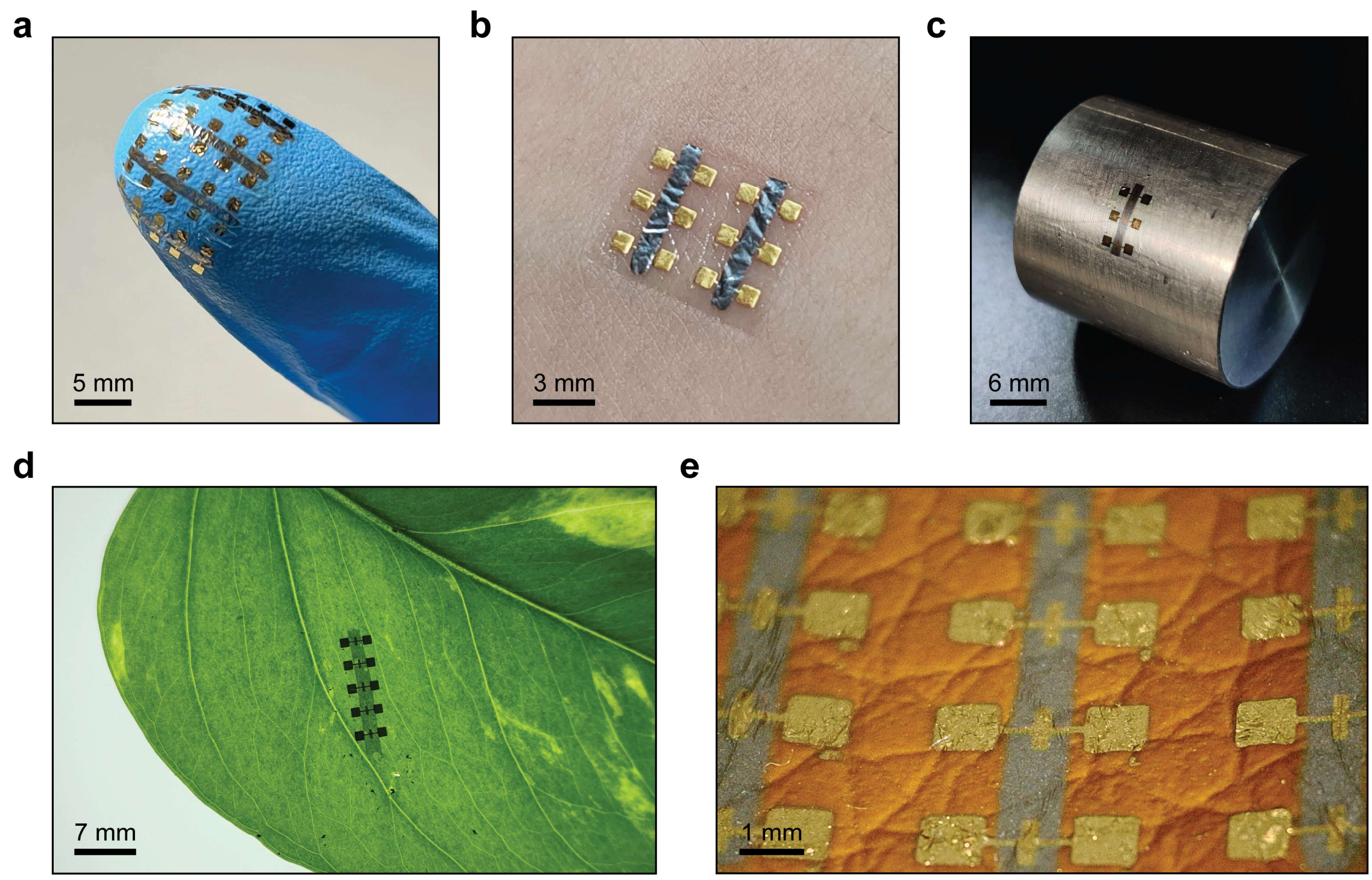

**Figure 3: Conformal integration of $MoS_2$-based devices on diverse substrates.** (a) Devices transferred onto a nitrile glove, illustrating adaptability to curved and textured surfaces. (b) Devices adhered to human skin, showing excellent conformability and adhesion. (c) Devices transferred onto a cylindrical metallic surface. (d) Devices successfully transferred onto a natural leaf, underscoring versatility for bio-integrated applications. (e) Close-up of the devices on synthetic leather, demonstrating compatibility with artificial materials.

Next, to demonstrate adaptability and conformal integration capability, we transferred the tattoo paper and waterslide decal paper-based devices onto a range of unconventional substrates with different surface textures, curvatures, and mechanical properties (see Figure 3). Figure 3a shows the full device array integration onto a nitrile glove, indicating the seamless conformity of the transferred film onto the soft, elastic, and low-friction surface. Figure 3b shows the physical adhesion of the film onto human skin, a biologically relevant surface known for its rough texture, curvature, and moisture content. The film maintained intimate contact with the skin without any visible delamination or cracking even under repeated stretching (see Video S3). Figure 3c shows the transfer on the curved metallic substrate, emphasizing the process's compatibility also with rigid, curved surfaces. The zoomed-in optical microscope image in Figure S8 shows that the device conforms well to the surface geometry, with no wrinkling, air gaps, or delamination. Figure 3d shows the integration of the devices onto a natural leaf, a fragile, moisture-containing, and irregular surface. Figure 3e highlights the successful adhesion of the device array onto synthetic polyurethane (PU) leather with a lychee-pattern surface that reproduces the grain and compliance of natural leather. Because of its pronounced surface roughness, soft mechanics, and complex topography (often exceeding those of human skin), we employed this PU leather in the following sections as an artificial testing platform for device characterization. This surrogate surface effectively captures key mechanical and topographical features of skin while providing a safe, reproducible, and standardized environment for evaluating transfer quality and device performance. Scanning electron microscopy (SEM) image presented in Figure S9 demonstrates the conformal adhesion of a tattoo device on the synthetic leather.

Although the adaptability of the devices to various platforms is illustrated using a single transfer medium in Figure 3, it should be emphasized that we do not observe any fundamental limitation for transferring the devices onto the demonstrated substrates using either transfer paper. Both transfer approaches are broadly applicable across these platforms and enable low-cost, high-throughput transfer of conformable devices.

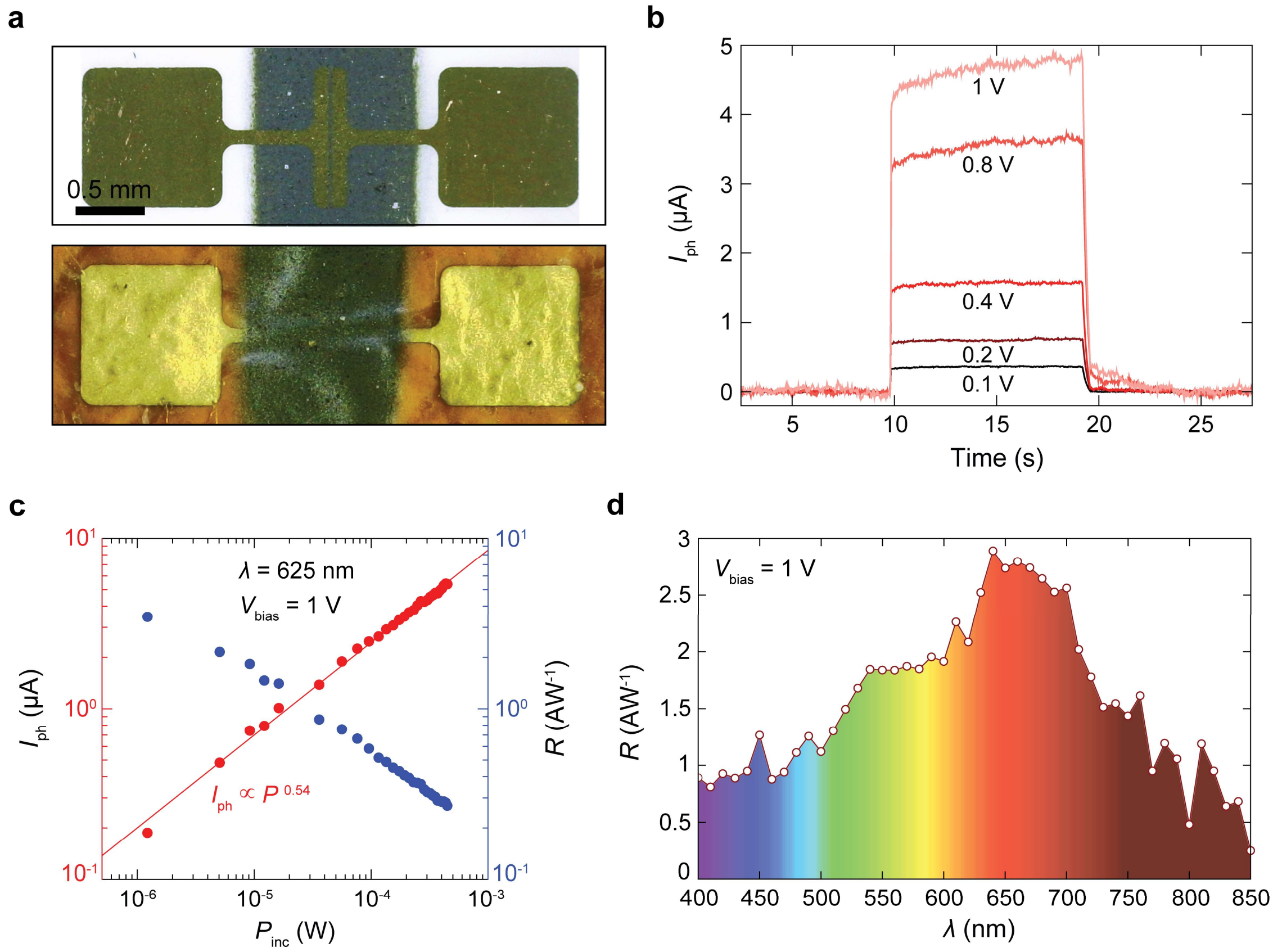

**Figure 4: Optoelectronic performance of $MoS_2$ photodetector fabricated on a waterslide decal paper after transfer onto synthetic leather.** (a) Optical images in the top and bottom panels show a single device before transfer (on waterslide decal paper) and after transfer onto synthetic leather, respectively. (b) Time-resolved photocurrent measurements under different bias voltages (625 nm, 0.45 mW), illustrating stability and reproducibility. (c) Log-log plot of photocurrent ($I_{ph}$) versus light power ($P_{inc}$) and corresponding responsivities. Photocurrent shows a non-linear dependence on illumination power with a power-law exponent of 0.54. (d) Spectral responsivity as a function of the light wavelength, highlighting the broadband sensitivity of the $MoS_2$ photodetector.

In the following sections, we systematically examine the performance of the devices after transferring them onto various platforms, such as synthetic leather and leaves, to evaluate their ability to operate reliably on substrates with different surface properties. A series of measurements was performed to characterize their sensitivity to light and temperature fluctuations, and to evaluate their performance when they are operated as field-effect transistors.

Figure 4 reveals the detailed optoelectronic performance of a waterslide decal paper-based device after its transfer onto synthetic leather. Optical images of a single device on waterslide decal paper and after transfer onto synthetic leather are shown in Figure 4a. Figure 4b presents the time-resolved photocurrent response measured under different bias voltages ($V_{bias}$), under illumination at a wavelength of 625 nm and a power of 0.45 mW. The device exhibits stable and reproducible switching between illuminated and dark states, indicating that the transferred device retains its photo-switching capability. The response time of the device was estimated as <40 ms, limited by the response time of the read-out electronics. Figure 4c shows the dependence of photocurrent ($I_{ph}$) on the incident light power, plotted on a log-log scale. The photocurrent exhibits a clear power-law dependence, following $I_{ph} \propto P^{\alpha}$, with a power exponent $\alpha = 0.54$.[38] This sublinear behavior is typical for photodetectors based on 2D materials, where processes such as trap state filling or recombination through defect states influence the carrier dynamics.[39] The corresponding responsivity ($R$) values can be calculated using the formula:

$$R = \frac{I_{ph}}{P_{inc}} \frac{A_{spot}}{A_{eff}}$$

where $I_{ph}$ is the measured photocurrent, $P_{inc}$ is the total incident light power, $A_{spot}$ is the total area of the illumination spot, and $A_{eff}$ is the effective device area, which is obtained by multiplying the total area of the device photoactive channel with a factor $c$. Here, $c$ is a correction factor ($0 < c \leq 1$) representing the fraction of the channel covered by $MoS_2$ flakes. This parameter was extracted from optical coverage analysis. The decreasing incident light intensity causes a gradual increase in the responsivity, which reaches a maximum value of 3.5 A $W^{-1}$ at the lowest illumination power of ~1 μW. The spectral responsivity of the device, shown in Figure 4d, confirms that the photodetector responds broadly across the visible spectrum (expected for multilayer $MoS_2$), with a peak responsivity in the 650-700 nm range, which is consistent with the direct band-to-band transition of $MoS_2$. To demonstrate the reproducibility and robustness of the method, several additional photodetector devices were transferred onto both synthetic leather and a curved metallic surface. Accordingly, 9 out of 10 devices remained functional and exhibited consistent photoresponse behavior after their transfer (see Figure S10). Furthermore, devices fabricated on two different transfer papers were also transferred onto leaves and operated as photodetectors, with the corresponding results presented in Figures S11–S14.

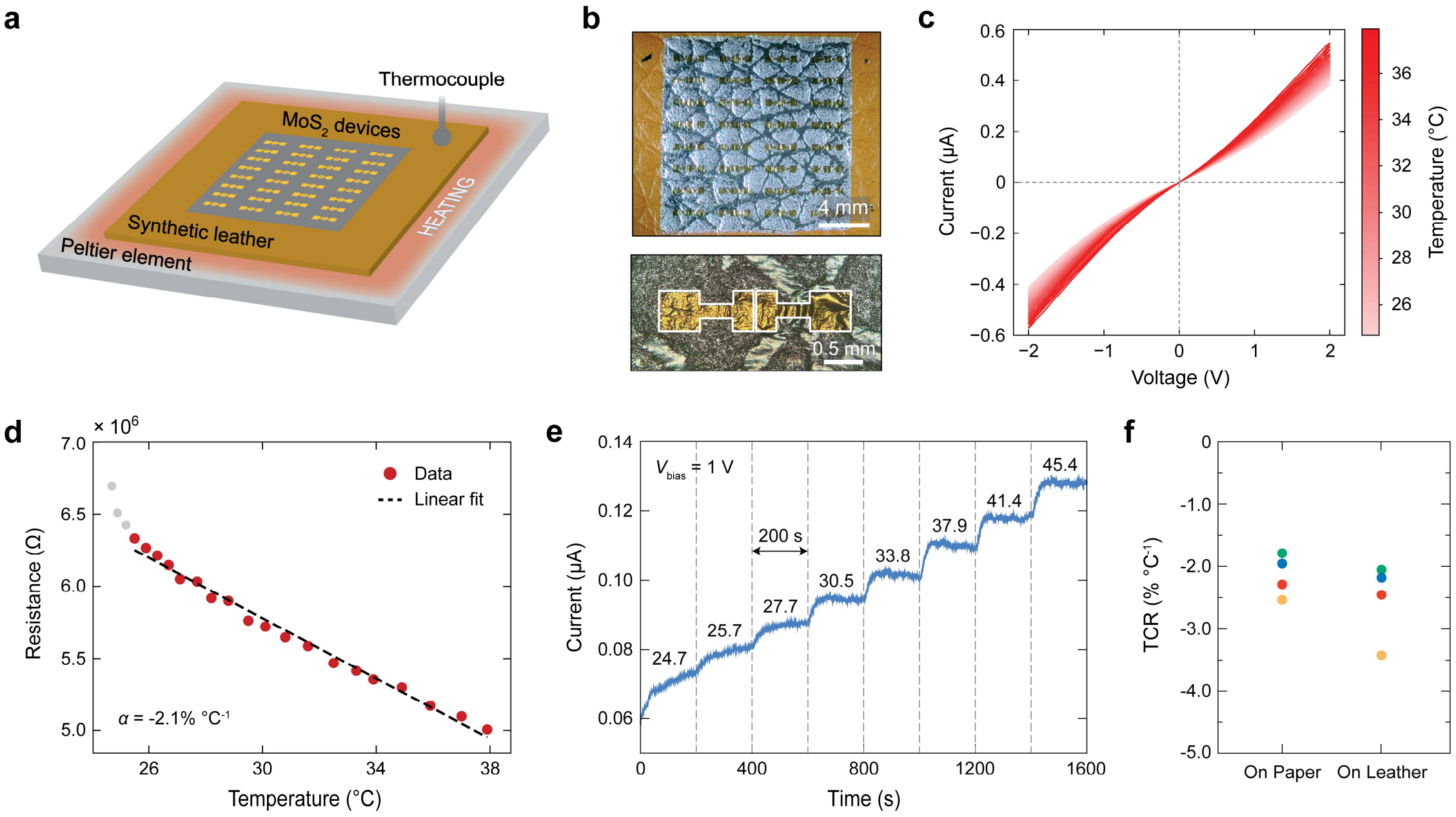

**Figure 5: Temperature-dependent electrical characterization of $MoS_2$-based tattoo devices.** (a) Simplified schematic of the experimental setup used for temperature-sensing measurements. (b) Optical images of the $MoS_2$ device array fabricated onto synthetic leather (top) and a magnified view of a single device (bottom). (c) Evolution of the *I*–*V* characteristics with increasing temperature, from room temperature up to 38 °C. (d) Temperature dependence of electrical resistance during heating. A high TCR value of −2.1 % °C$^{-1}$ was extracted from the linear fit (black dashed line). Grey dots correspond to the data excluded in the linear fit. (e) Real-time monitoring of the current response during stepwise heating. The numbered points within the plot correspond to the temperature values recorded once the system reached thermal equilibrium at each step. (f) Comparison of TCR values from four devices, obtained before transfer (on tattoo paper) and after transfer (on synthetic leather).

Next, the devices were operated as thermistors, where their resistance variation with temperature was monitored to evaluate the thermal sensitivity. For thermistor applications, tattoo paper was preferred due to its better conformability compared to waterslide decal paper. Improved conformal contact between the device and the target surface enhances heat transfer, enabling faster thermal equilibration and more accurate temperature sensing. After transferring $MoS_2$ films onto tattoo paper, Au electrodes were deposited through a custom-designed shadow mask via thermal evaporation[40].

This custom-made mask enables the fabrication of 32 individual devices within a compact 1.5 $cm^2$ area, allowing high-throughput characterization under identical fabrication and measurement conditions. The completed films were subsequently transferred onto synthetic leather using the scooping technique.

For temperature-dependent electrical characterization, the $MoS_2$-based tattoo devices were placed on a Peltier element, which was used as a controllable heating platform. The experimental setup is schematically illustrated in Figure 5a, and real images of the prepared sample are given in Figure 5b. The Peltier element was operated in reverse-polarity configuration by a DC power supply (TENMA 72-2715) to induce heating. The applied voltage to the Peltier element was gradually increased in discrete steps to modulate the substrate temperature. At each temperature point, current–voltage (*IV*) measurements were performed to monitor the device response to the increasing temperature. Note that, after each temperature increment, a 30-second stabilization period was allowed to ensure thermal equilibrium before recording the *IV* characteristics. A thermocouple was positioned in close proximity to the device region on top of the synthetic leather surface to accurately record the local temperature during measurements.

Figure 5c clearly illustrates the impact of temperature on the *I-V* characteristics, showing that higher temperatures lead to increased current levels, which are more likely associated with a combination of thermally activated charge carriers, enhanced carrier injection from the contacts and thermally activated hopping between interconnected $MoS_2$ nanosheets, originating from reduced junction resistance between adjacent flakes. Due to the supralinear *I-V* characteristics, which are primarily attributed to bias-induced

Joule self-heating of the ultrathin $MoS_2$ tattoo thermistors, the resistance values were extracted by linear fitting within a small voltage range where the curves are approximately linear. Figure 5d presents the resistance variation as the temperature was swept in the range of 25 °C to 38 °C, which is typically the range used for body-temperature sensors and wearable thermistors. The device resistance exhibited a linearly decreasing trend with increasing temperature, indicating negative temperature coefficient (NTC) thermistor behavior of semiconducting $MoS_2$.[41]

Temperature coefficient of resistance (TCR), sometimes referred to as sensitivity, is a key parameter to quantitatively describe the sensitivity of the device's resistance to temperature and can be expressed as:

$$\mathrm{TCR} = \frac{1}{R_{ref}} \frac{dR}{dT}$$

where $R_{\mathrm{ref}}$ is the reference resistance, measured at a temperature closest to 36.5 °C, which is within the typical range of normal human body temperature. By performing a linear fit, the TCR value was determined to be −2.1 % $°C^{-1}$. This TCR value is higher than previously reported $MoS_2$-based temperature sensors as given in Table S1, which also summarizes previously reported TCR values for temperature sensors based on different sensing materials implemented on flexible and stretchable platforms. Figure 5e presents the real-time current response of the $MoS_2$ tattoo thermistor during consecutive heating, with the temperature swept between 24.7 and 45.4 °C. After each temperature increment, a 200 s holding period was applied to perform the read-out before jogging to the next temperature value. The current exhibited a rapid, stepwise increase with rising temperature, maintaining a stable level at each temperature plateau. Figure 5f summarizes the TCR values extracted from four $MoS_2$-based tattoo devices, both before transfer (on tattoo paper) and after transfer onto synthetic leather. Across both substrates, the devices demonstrated TCR values of comparable magnitude, indicating uniform sensing characteristics. The temperature-dependent resistance of additional devices following transfer onto synthetic leather is shown in Figure S15.

We further explore the use of the $MoS_2$ conformal devices as proof-of-concept FETs. To build top-gated $MoS_2$-based tattoo FETs, we used a commercially available ionic conductive hydrogel (see Materials and Methods), which was placed on top of the tattoo paper devices transferred onto synthetic leather to serve as an electrolyte−gate dielectric. The ionic gel, a composite of an ionic liquid and a flexible polymer matrix, exhibits electrochemical behavior similar to that of the ionic liquid while offering greater mechanical robustness and solid-like characteristics. When a gate voltage is applied, the mobile positive and negative ions within the gel migrate toward the gate electrode and $MoS_2$ interfaces, respectively, forming two electrical double layers, one at each interface. The electrical double layer at the gel/semiconductor interface acts as an ultrathin gate dielectric, creating a large interfacial capacitance and enabling efficient tunability in channel conductance within a small gate voltage range.

It is important to note that present devices are a proof-of-concept demonstration; we did not specifically engineer the ionic gate, and the current implementation employs a relatively thick ionic gel. Nevertheless, we are confident that the device architecture can be significantly optimized in future studies by research groups with expertise in ionic-gating materials.

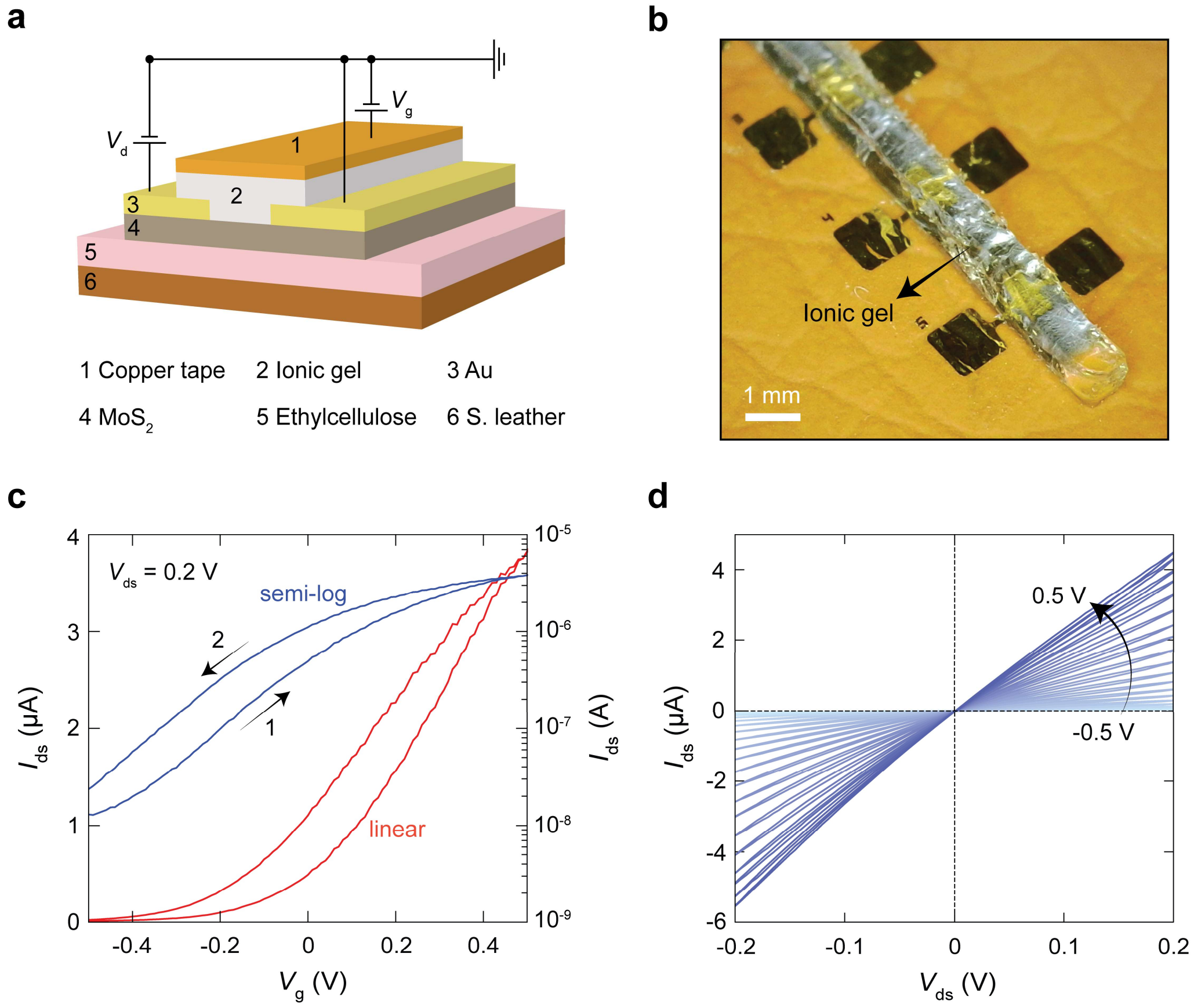

**Figure 6: Ionic gel-gated $MoS_2$ tattoo-FET on synthetic leather.** (a) 3D cross-sectional schematic illustrating the layer-by-layer FET structure. (b) Optical images of an ionic-gel-transferred tattoo device on synthetic leather. (c) Semi-log and linear-scale plot of transfer characteristics under forward (1) and backward (2) gate voltage sweep for a $V_{ds}$ of 0.2 V. (d) Output characteristic curves under gate voltage sweeps from –0.5 V to 0.5 V.

Tattoo FETs were obtained by attaching a thin, strip-shaped cut piece of ionic gel onto a single-line device array transferred onto synthetic leather, selectively covering the channel regions to ensure proper gating while minimizing contact with the source and drain electrodes. Later, a copper tape was placed on top of the gel to serve as the gate electrode. Figures 6a and 6b show the device composition and an optical image of the devices after ionic gel transfer (prior to attaching the copper tape gate electrode on top), respectively. Figure 6c shows the transfer characteristics for one of the fabricated tattoo FETs in semi-log and linear scale, measured at a drain–source voltage ($V_{ds}$) of 0.2 V. Due to the slow ion migration within the ionic gel under the applied gate voltage, the gate voltage was swept in 0.01 V steps every 3 s. The device exhibits typical n-type transistor behavior and can be operated within a small gate voltage range, with a threshold voltage ($V_{th}$) of ~10 mV, which was extracted by linear extrapolation of the linear-scale $I_{ds} - V_g$ curve to the $V_g$-axis. The device shows an on/off ratio ($I_{on}/I_{off}$) of $3\times10^2$, with a subthreshold slope of 254 mV $dec^{-1}$. The effective field-effect mobility (μ) can be estimated from the equation,

$$\mu = g_m \frac{L}{W_{eff} C_G V_{ds}}$$

where $g_m = (\partial I_{ds}/\partial V_g)$ is the transconductance, $V_{ds}$ is the drain-source voltage, $C_G$ is the gate capacitance of the ionic gel per unit area, and $L$ and $W_{eff}$ are the channel length and effective channel width, respectively. $W_{eff}$ was determined by multiplying the total channel width, $W$, by a flake coverage factor of $c$, which is 0.8 for the corresponding device. The capacitance of the ionic gel was measured as $9.6\times10^{-3}$ F/m$^2$ using RC circuit measurements, which will be reported in a forthcoming study. This measured areal capacitance is consistent with values commonly reported for ionic gels and ionic liquids.[42–45] $\mu$ was estimated using the maximum transconductance value extracted from the transfer curve and was obtained to be 1.2 cm$^2$ V$^{-1}$ s$^{-1}$ for this device. Figure 6d shows the gate-dependent *IV* curves of the tattoo FET, where the gate voltage was swept from –0.5 V to 0.5 V. The more linear, ohmic behavior observed under ionic gel gating is likely associated with electrostatic carrier accumulation induced by the ionic gel near the $MoS_2$/electrode interface. The increased local carrier density can enhance carrier injection from the contacts and reduce the effective contact resistance. In addition, ionic-gel-induced electrostatic doping of the $MoS_2$ channel may further contribute to the enhanced device conductivity. It should also be noted that the FET measurements were performed within a significantly smaller $V_{ds}$ compared to the photodetector and thermistor measurements in order to minimize ionic leakage currents and avoid undesired electrochemical reactions within the ionic gel. This reduced operating bias additionally suppresses self-heating effects and contributes to the more linear current–voltage characteristics observed in the FET devices.

To assess the device-to-device variability and reproducibility of the method, we have characterized 6 FETs in total under ionic gel gating on synthetic leather. Key performance parameters for all devices are summarized in Table S2, demonstrating that their overall electrical characteristics are closely aligned with only modest variation. The extracted mobilities from the transfer curves (see Figure S16) span from a minimum of 0.23 $cm^2 V^{-1} s^{-1}$ to a maximum of 17.6 $cm^2 V^{-1} s^{-1}$, yielding an average mobility of 4.55 $cm^2 V^{-1} s^{-1}$. Narrow gate window operation yields extremely small $V_{th}$ values with minimal variation, ranging from –41 mV to 10 mV (see Figure S17), which is important for the development of flexible and wearable platforms that operate with low energy consumption. In our previous study on roll-to-roll mechanically exfoliated $MoS_2$ films, the highest measured mobility was 1.36 $cm^2 V^{-1} s^{-1}$. However, those devices employed a conventional back-gate $SiO_2$ configuration, where the relatively low gate capacitance limits electrostatic modulation of the channel conductivity. In contrast, the significantly higher capacitance of ionic gel dielectrics enables much stronger gate coupling and more efficient carrier modulation, resulting in substantially enhanced mobility values. It should be noted that the mobility values reported here for the ionic gel–gated devices are among the highest reported for $MoS_2$ nanosheet-network-based FETs.[42,46–50]

As an alternative to ionic gel gating, ethylcellulose layers can serve as a gate dielectric to modulate the channel conductivity of devices on tattoo paper.[28] However, the modulation is weaker, likely due to the lower capacitance of ethylcellulose layer ($5.3\times10^{-5}$ F/m$^2$) and the increased dielectric thickness arising from the double-layer ethylcellulose structure present in this device configuration needed to avoid gate leakage. The reader is referred to Figure S18 for further details on the device architecture and FET characterization.

Even when a uniform areal coverage is achieved through iterative film transfer, devices based on flake-network films can still exhibit significant variation in their electronic properties. This variation is primarily associated with the percolative nature of charge transport through the interflake junction network. In films produced by methods such as roll-to-roll mechanical exfoliation, electrochemical exfoliation, and liquid-phase exfoliation, current flows through percolation pathways defined by flake–flake contacts. The resistance of these pathways is largely governed by factors such as the overlap area between flakes, their relative alignment, and the lateral size of the flakes. When percolation pathways consist of well-aligned flakes with large overlap areas, or when they are formed by larger flakes that reduce the number of junctions along the conduction path, the overall device resistance is lowered. However, because these films are composed of randomly distributed flakes (each transfer introduces additional randomness into the network morphology), the resulting percolation network can vary significantly from device to device. Consequently, some devices may be dominated by highly resistive junctions, while others benefit from more conductive pathways. This intrinsic variability in the percolation network is a key factor leading to fluctuations in the measured mobility and overall electronic performance.

The device performance could be further improved by achieving better control over nanosheet size distribution by optimizing exfoliation parameters (e.g., pressure and speed) or improving deposition processes to obtain better network uniformity.[35,42] In combination with post-processing techniques such as thermal annealing, compression/calendering, or covalent functionalization of the nanosheet network, device-to-device variations could be further reduced through improved interflake coupling and an enhanced degree of nanosheet alignment, leading to more uniform percolative charge transport across the large-area films.[18,51–53]

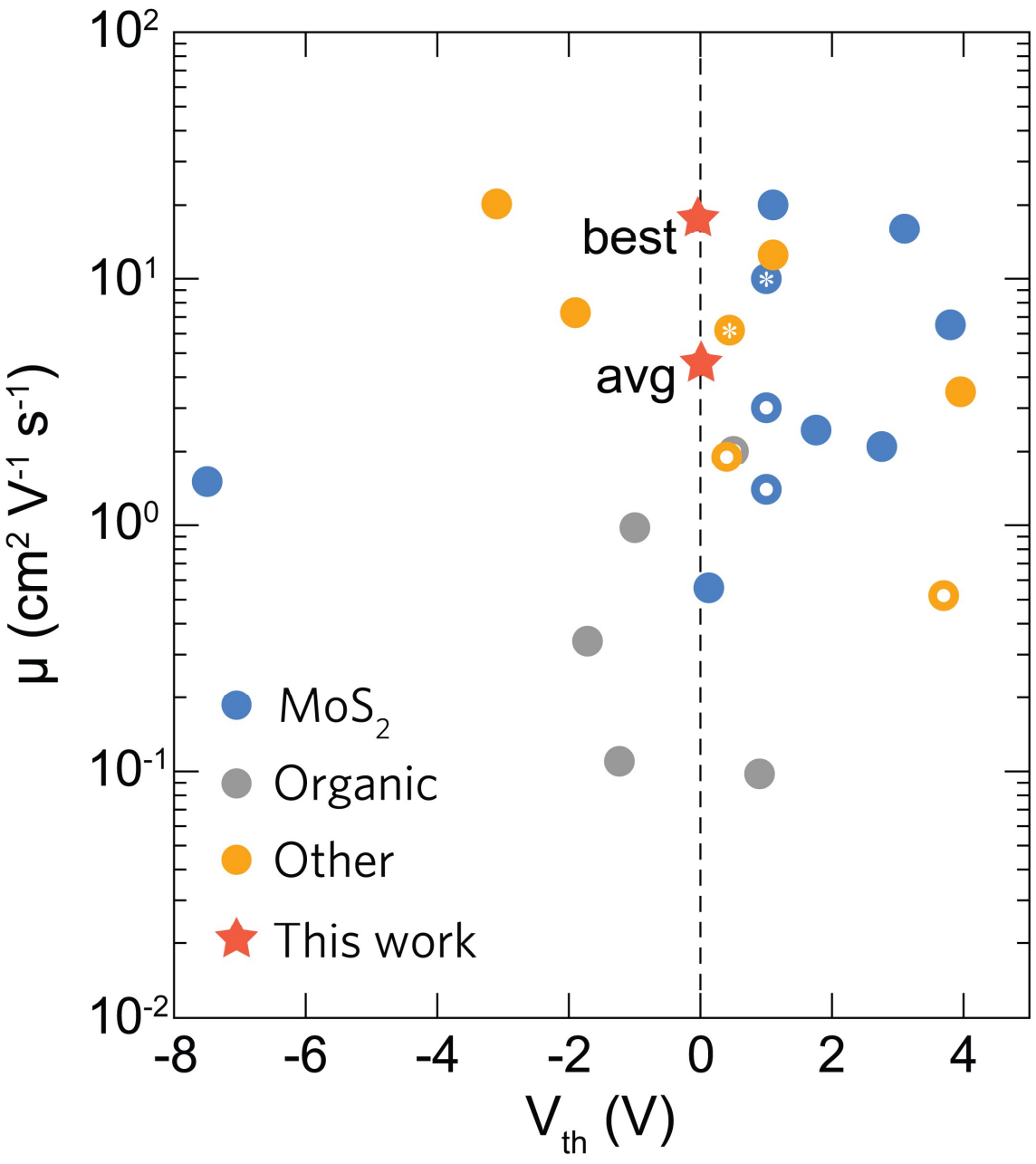

**Figure 7: Benchmarking across flexible and stretchable FETs.** The plot shows the mobility ($\mu$) as a function of threshold voltage ($V_{th}$), comparing the average and highest values from our devices with those reported for transistors fabricated on flexible or stretchable substrates. For clarity, the devices are classified according to their active channel materials: $MoS_2$ (blue), organic polymers (gray), and other inorganic and carbon-based materials (orange). Open symbols denote ionically gated devices. Data points marked with an asterisk (*) indicate studies in which the $V_{th}$ was not reported and was instead extracted by us from the transfer characteristics provided in the original publications.

Figure 7 presents reported $\mu$ values for transistors fabricated on flexible or stretchable platforms as a function of $V_{th}$, including our average and best values for direct comparison. The devices were classified into groups such as $MoS_2$, organic polymers, and other inorganic and carbon-based materials, to separate materials exhibiting fundamentally different charge-transport mechanisms. Additionally, ionically gated transistors are represented with open symbols to distinguish them from devices gated with solid-state dielectrics, which operate based on fundamentally different principles. It is worth noting that our values were achieved without any post-fabrication treatments to improve device performance, such as acid treatments or high-temperature annealing, and without performing measurements under vacuum or encapsulated conditions. Accordingly, our $\mu$ values are significantly higher than those reported for organic field-effect transistors and also exceed those of ionically gated networks. The average $\mu$ value we obtained is on the same order of magnitude as the highest values reported in the literature.[54–56] Moreover, the $V_{th}$ values of our devices are among the lowest reported for comparable flexible and stretchable FET platforms. A more comprehensive summary of the reported FET parameters for each study included in Figure 7 is provided in Table S3.

## CONCLUSIONS

In this work, we demonstrate that combining high-throughput roll-to-roll mechanical exfoliation of $MoS_2$ with commercially available temporary tattoo and waterslide decal transfer papers provides a low-cost and scalable route for fabricating conformal devices. Unlike liquid-phase exfoliation followed by printing (where residual solvents, poor interflake connectivity, and limited film uniformity often compromise electronic performance) and unlike CVD-grown films (which require expensive infrastructure and complex transfer processes for the as-grown films), roll-to-roll mechanical exfoliation provides dry, interconnected $MoS_2$ films that can be readily integrated into ultrathin, transferable platforms. By leveraging this material platform together with simple decal-based transfer strategies, we fabricate ultraconformal devices that operate reliably on rough and compliant surfaces. The resulting tattoo-based photodetectors, thermistors, and ionic-gel-gated transistors exhibit high responsivity, large temperature coefficients of resistance, and low-voltage, high-mobility transistor operation.

## MATERIALS AND METHODS

### Device fabrication

$MoS_2$-based tattoo devices were fabricated on temporary tattoo papers and waterslide decal papers, which were supplied from TheMagicTouch Spain S.L.

(www.themagictouch.es) and Hayes Paper Co. (www.hayespaper.com), respectively. The $MoS_2$ thin films were obtained through the roll-to-roll mechanical exfoliation of a natural molybdenite mineral (Molly Hill Mine, Quebec, Canada). This system is based on continuous exfoliation of bulk van der Waals crystals through two rolling cylinders in contact, wrapped with adhesive Nitto tape, enabling peeling of crystal layers as the cylinders rotate. The roll-to-roll exfoliation process was carried out for 1 min.

After exfoliation, bar-shaped $MoS_2$ films were patterned on the transfer papers by successively transferring flakes from Nitto tape onto tattoo paper or waterslide decal paper using a bar-shaped stencil mask. The stencil mask was fabricated from a 100 μm thick Mylar sheet using a smart cutting machine (Cricut Maker 3). The dimensions and spacing of the bar-shaped openings were designed to match the geometry and column-to-column spacing of the 4 × 5 electrode array in the commercial shadow mask (Ossila) used for electrode deposition, thereby enabling accurate alignment of the electrode channels with the films during mask placement.

After attaching the flake-containing Nitto tape to the tattoo paper (waterslide decal paper), the samples were annealed on a hot plate at 70 °C (100 °C) for 5 min to induce thermal release of the flakes. The narrow openings of the stencil mask can hinder proper contact between the tape and the substrate. To ensure complete attachment, gentle pressure was applied over the openings using a cotton swab shortly after placing the sample on the hot plate, once the substrate had begun to warm.

After annealing, the Nitto tape was removed immediately upon removing the sample from the hot plate, while still warm. This is particularly important to prevent tearing of the transferable polymer films on the paper. The transfer process was repeated to form a continuous, interconnected film of overlapping flakes. Finally, source–drain electrodes were obtained by depositing 80 nm thick Au contacts via thermal evaporation using commercial (Ossila) or home-built shadow masks.

For FET characterization, a commercially available conductive hydrogel (C100AE EMS hydrogel pads, model JP-EMS-11) of the type used in TENS (transcutaneous electrical nerve stimulation) pads was transferred onto the semiconducting $MoS_2$ channel to form an ionic interface layer. These hydrogels consist of water-rich polymer networks (typically PVA based water gels) containing mobile ions (dissolved salts), providing stable ionic conductivity and conformal contact with soft surfaces. To obtain the gate contact, a copper tape was attached on top of the ionic gel.

**Characterization**

The atomic force microscopy images were taken in dynamic mode with a resonance frequency of 76 kHz, using a cantilever oscillation amplitude of 1.7 V. The tip used was a PPP-FMR-50 from nanosensors and the analysis was done with a commercial AFM from Nanotec.

Field-emission scanning electron microscopy images were collected on an FEI Nova NanoSEM 230 scanning electron microscope with a Schottky field-emission gun equipped with a W ion source, using aluminum as a support.

For Raman measurements, transferable hydrophobic polymer films from tattoo paper and waterslide decal paper were deposited onto $SiO_2$/Si substrates and annealed at 80 °C for 15 min. Raman spectra of the transferred films were then acquired under ambient conditions using a confocal Raman microscope (MonoVista CRS+, Spectroscopy & Imaging GmbH) with 532 nm excitation from a continuous wave (CW) solid-state laser. A 300 lines/mm diffraction grating was used, providing a spectral resolution of 6 $cm^{-1}$.

The incident light power was set to 0.54 mW and focused through a 50× objective (NA = 0.75), providing a spot size of 2 μm.

**Electrical Measurements**

All electrical measurements were performed under atmospheric pressure at room temperature using a home-built probe station. In the FET measurements, the source–drain terminals were biased using probes connected to a Keithley 2450 source-meter unit, while a controlled gate-voltage sweep was applied to the gate terminal through a separate probe using two programmable benchtop power supplies (Tenma 72-2715) connected in back-to-back configuration.

**Optoelectronic Measurements**

Photocurrent measurements under varying power and bias conditions were performed using a fiber-coupled LED source at a wavelength of 625 nm (Thorlabs M625F2), operated by a LED driver (Thorlabs LEDD1B)[57]. The LED output power was varied by tuning the LED drive current with a Tenma power supply unit (model 72-2715). Wavelength-dependent photocurrent measurements were conducted using a tunable xenon lamp source (Bentham TLS120Xe), sweeping the incident light across the visible range from 400 nm to 850 nm in 10 nm increments.

ASSOCIATED CONTENT

**Supporting Information**

Supplementary Information includes:

**Figure S1.** Raman spectra of transferable polymer films of tattoo and waterslide decal papers

**Figure S15.** Temperature-dependent resistance of devices transferred onto synthetic leather

**Table S1.** Literature summary of reported temperature sensors

**Table S2.** Key device performance metrics of ionic gel-gated $MoS_2$ tattoo FETs on synthetic leather

**Figure S16.** Transfer characteristics of ionic gel gated $MoS_2$ tattoo FETs

**Figure S17.** Linear-scale transfer characteristics of ionic gel–gated $MoS_2$ tattoo FETs

**Figure S18.** Gating tattoo devices through ethylcellulose

**Table S3.** Literature summary for FETs demonstrated on flexible and stretchable platforms

Supplementary Video Descriptions

## AUTHOR INFORMATION

### Corresponding Authors

Yigit Sozen yigit.sozen@csic.es, Andres Castellanos-Gomez andres.castellanos@csic.es

### Author Contributions

Y.S. lead data curation; formal analysis; investigation; methodology; the original draft writing. E.Z supported data curation, formal analysis, investigation, and methodology. J.J.R supported resources, supervision and methodology. A.C.-G. lead conceptualization, funding acquisition, project administration, resources, supervision, supported the methodology, and original draft writing. All authors contributed equally to review and editing.

**Conflict of Interest**

The authors declare that they have no known competing financial interests or personal relationships that could influence the work reported in this manuscript.

**Data Availability**

The datasets generated and/or analyzed during the current study will be made publicly available in the Zenodo repository through our community page (2D Foundry community at Zenodo) at: https://zenodo.org/communities/2dfoundry.

All relevant data supporting the findings of this study will be accessible upon publication.

**ACKNOWLEDGEMENTS**

The authors thank Dr. Carmen Munuera (ICMM-CSIC) for her support with the AFM measurements and useful discussions along the work. This work was funded by the Ministry of Science and Innovation (Spain) through the projects PRE2021-098348 and PID2023-151946OB-I00 and funded by the European Commission – NextGenerationEU (Regulation EU 2020/2094), through CSIC's Quantum Technologies Platform (QTEP). The authors also acknowledge funding from the European Research Council (ERC) through the ERC-PoC 2024, StEnSo project (grant agreement no. 101185235), and ERC-2024 SyG SKIN2DTRONICS (grant agreement 101167218). J.J.R. hired under the Generation D initiative, promoted by Red.es, an organisation attached to the Ministry for Digital Transformation and the Civil Service, for the attraction and retention of talent through grants and training contracts, financed by the Recovery, Transformation and Resilience Plan through the European Union's Next

Generation funds. ICMM-CSIC authors also acknowledge support from the Severo Ochoa Centres of Excellence program through Grant CEX2024-001445-S, funded by MICIU/AEI/10.13039/501100011033.

# Supporting Information:

# Scalable conformal electronics based on roll-to-roll exfoliated van der Waals semiconductors

*Yigit Sozen[1]*, Esteban Zamora-Amo[1], Juan J. Riquelme[1], Andres Castellanos-Gomez[1]**

*[1]2D Foundry Research Group. Instituto de Ciencia de Materiales de Madrid (ICMM-CSIC), Madrid, E-28049, Spain.*

*corresponding authors yigit.sozen@csic.es, andres.castellanos@csic.es

## Raman spectra of transferable polymer films of tattoo and waterslide decal papers

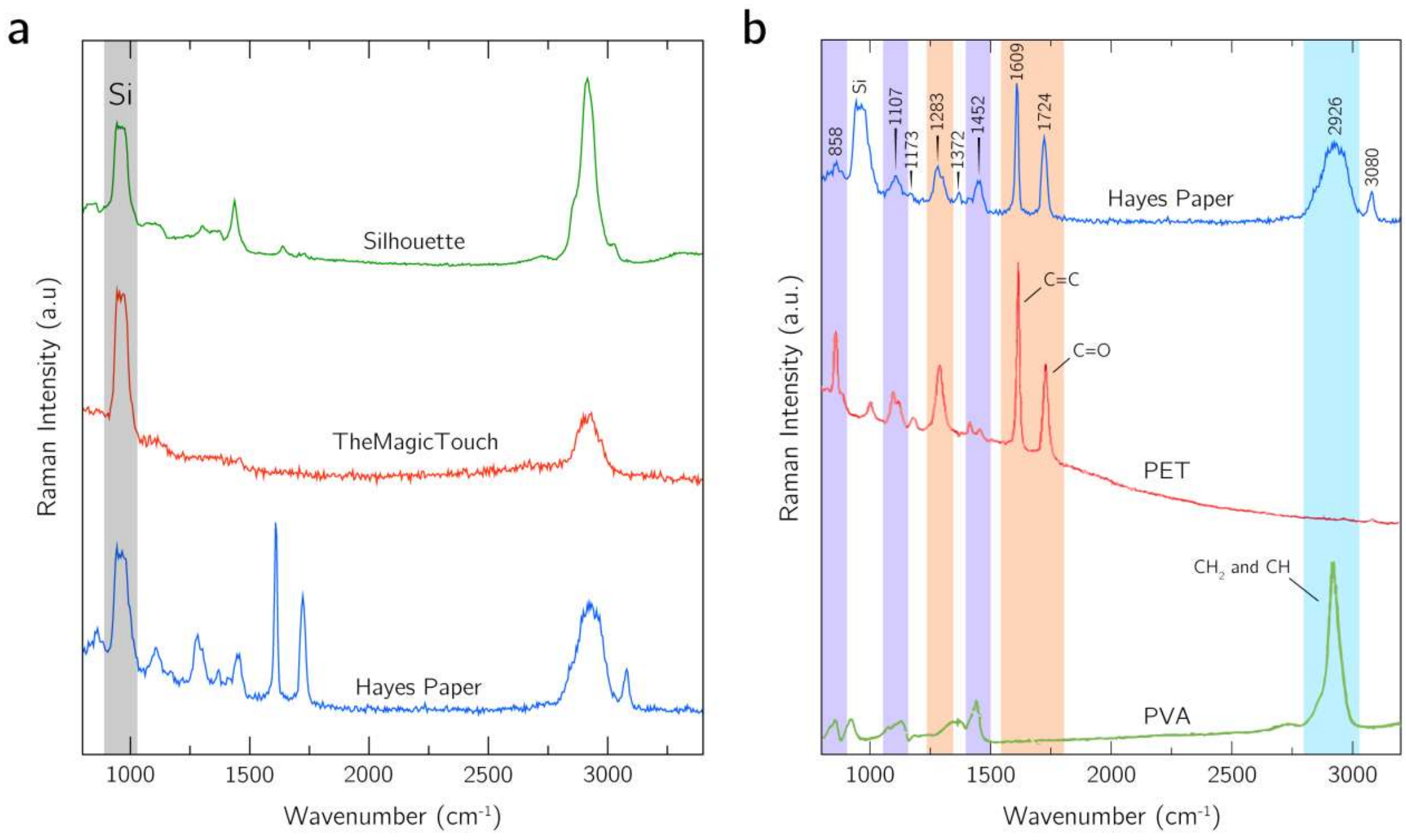

**Figure S1. Raman spectra of Silhouette, TheMagicTouch, and Hayes papers.** (a) Raman spectra acquired after transferring the top polymer films from Silhouette (green line), TheMagicTouch (red line), and Hayes papers (blue line) onto a $SiO_2$/Si (290 nm) substrate. Silhouette, another PVA-based temporary tattoo paper, is included to highlight the similarity of its Raman features with those of Hayes paper and served as a reference to elucidate the material composition of Hayes paper. The grey shaded area indicates the region where the second-order Raman peak of silicon is located. (b) The figure presents a direct comparison of the Raman spectrum of Hayes paper (blue line) with PET[1] (red line) and PVA[2,3] (green line) reported in previous studies. The blue and orange shaded regions indicate the spectral ranges in which the PVA and PET samples share common Raman peaks with Hayes paper, respectively. The violet regions, in particular, highlight the intervals where all spectra display similar Raman peaks. The intense peaks located at around 1609 $cm^{-1}$ and 1724 $cm^{-1}$ originate from C=C aromatic stretching and C=O stretching vibrations, respectively, which are visible in PET and Hayes. The broad peak around 2926 $cm^{-1}$ originates from the symmetric and asymmetric stretching vibrations in $CH_2$ and CH and is only apparent in PVA and Hayes paper. On the other hand, PET and Hayes paper share another peak located at 1283 $cm^{-1}$, which arises from C(O)−O stretching. The Raman signatures of PET and PVA confirm their presence within the structure of Hayes decal paper. The peak at 3080 $cm^{-1}$ can be attributed to impurities or the presence of other chemical species.

## Atomic force microscopy (AFM) analysis of the surface morphology and thickness of ethylcellulose and cross-linked PVA + PET films

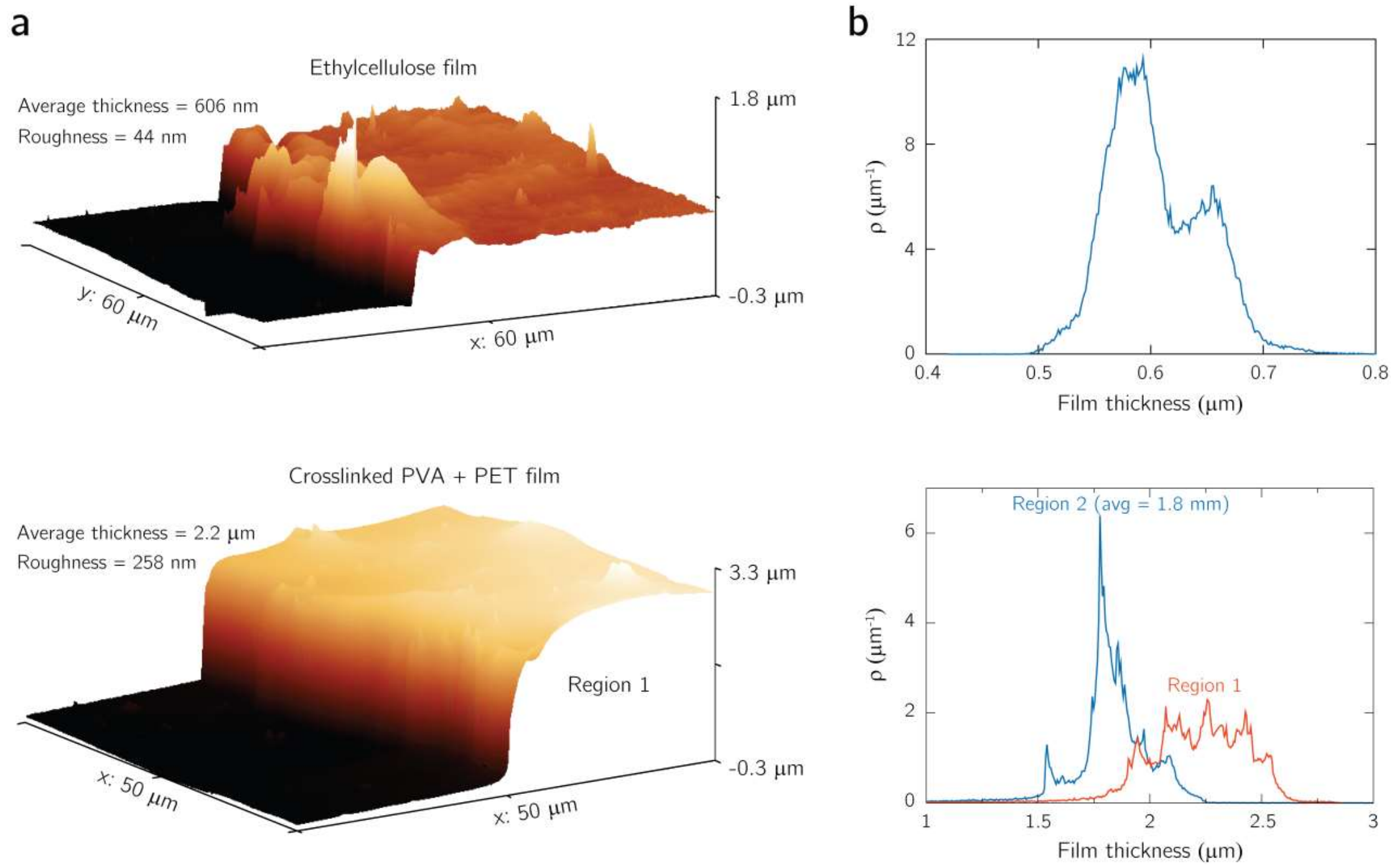

**Figure S2. AFM topography images of transferable polymer layers from TheMagicTouch 2.1 Tattoo paper and Hayes waterslide decal papers.** (a) Three-dimensional AFM images showing the surface morphology of the ethylcellulose film from TheMagicTouch (top) and the cross-linked PVA + PET film from Hayes paper (bottom) after transfer onto a $SiO_2$/Si substrate. To expose the underlying substrate, scratches approximately 5 µm wide were made on the films using a vinyl record player stylus. (b) Height profiles of the ethylcellulose (top) and crosslinked PVA (bottom) films extracted from the corresponding AFM images in (a). The height profile indicated as Region 2 for the cross-linked PVA film was obtained from a separate AFM image to illustrate the variation in film thickness. Three-dimensional AFM images were obtained using Gwyddion software[4].

## Evolution of the $MoS_2$ film with successive transfers on tattoo paper (TheMagicTouch Tattoo 2.1) and waterslide decal paper (Hayes)

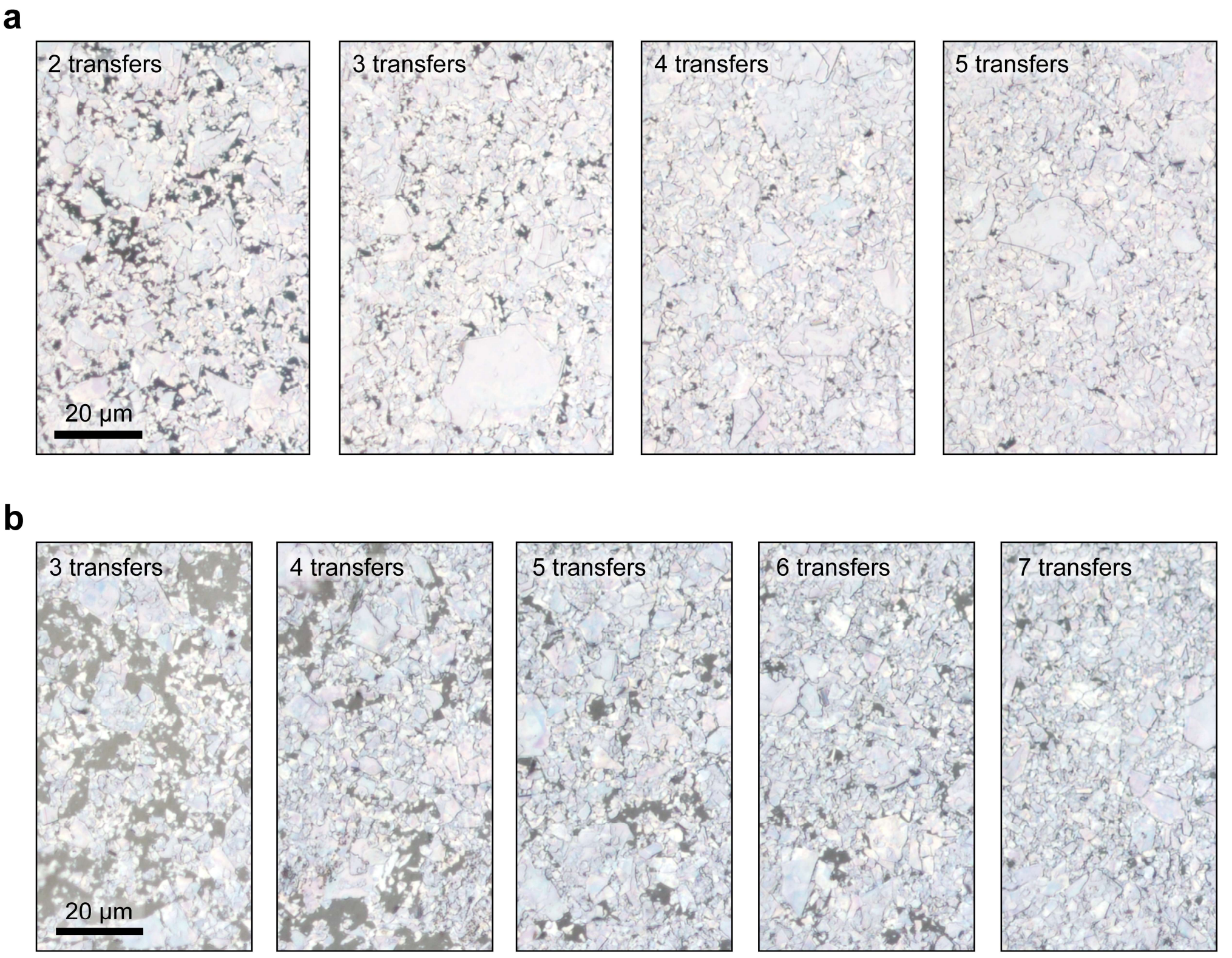

**Figure S3. $MoS_2$ nanosheet network formation through successive transfer cycles.** Optical microscope images showing the evolution of the $MoS_2$ film after each transfer step on (a) tattoo paper and (b) waterslide decal paper.

## Substrate coverage as a function of the number of transfer steps

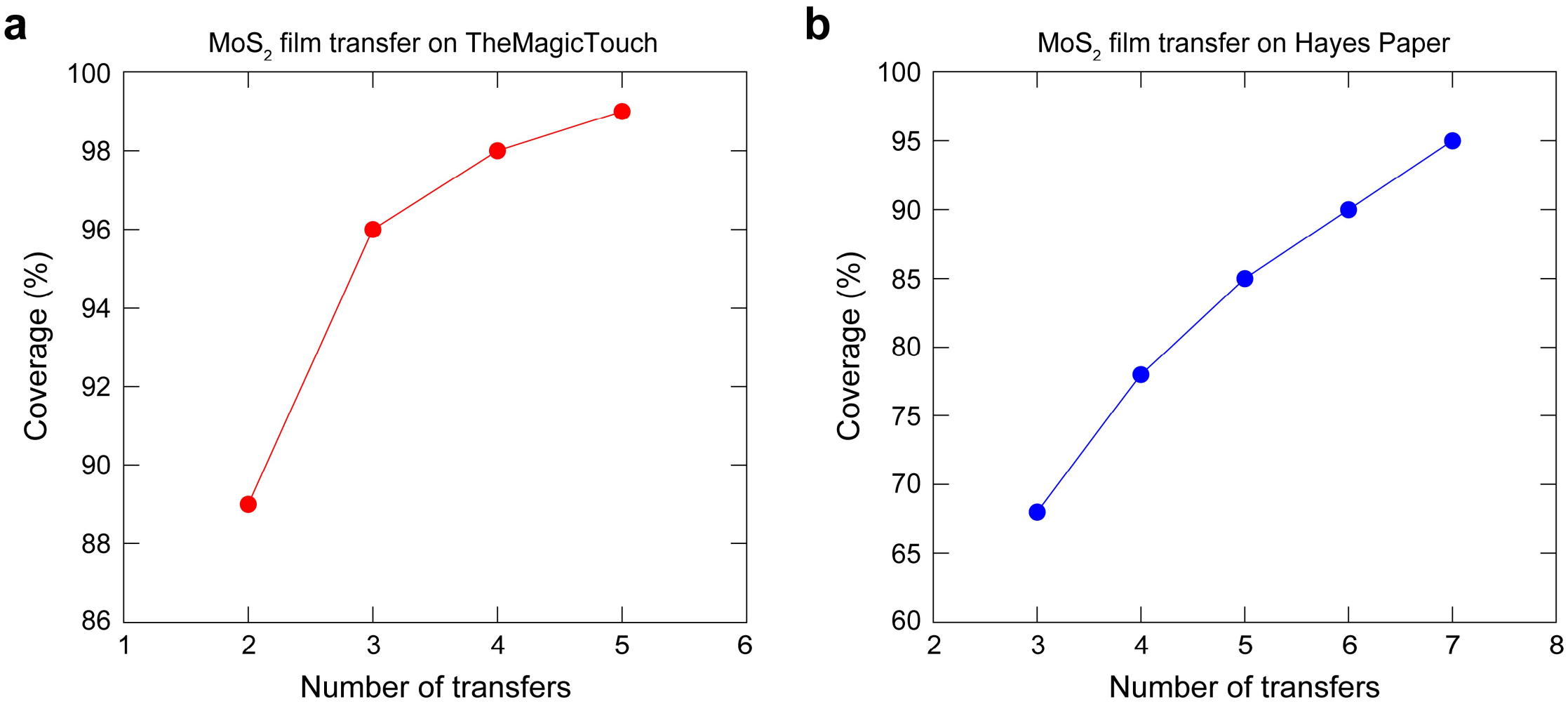

**Figure S4.** Evolution of substrate coverage (%) as a function of the number of transfer steps for (a) tattoo paper and (b) waterslide decal paper. The coverage values were obtained from optical images shown in Figure S3 using Gwyddion software[4].

## Atomic force microscopy (AFM) characterization of $MoS_2$ flakes on tattoo paper and waterslide decal paper

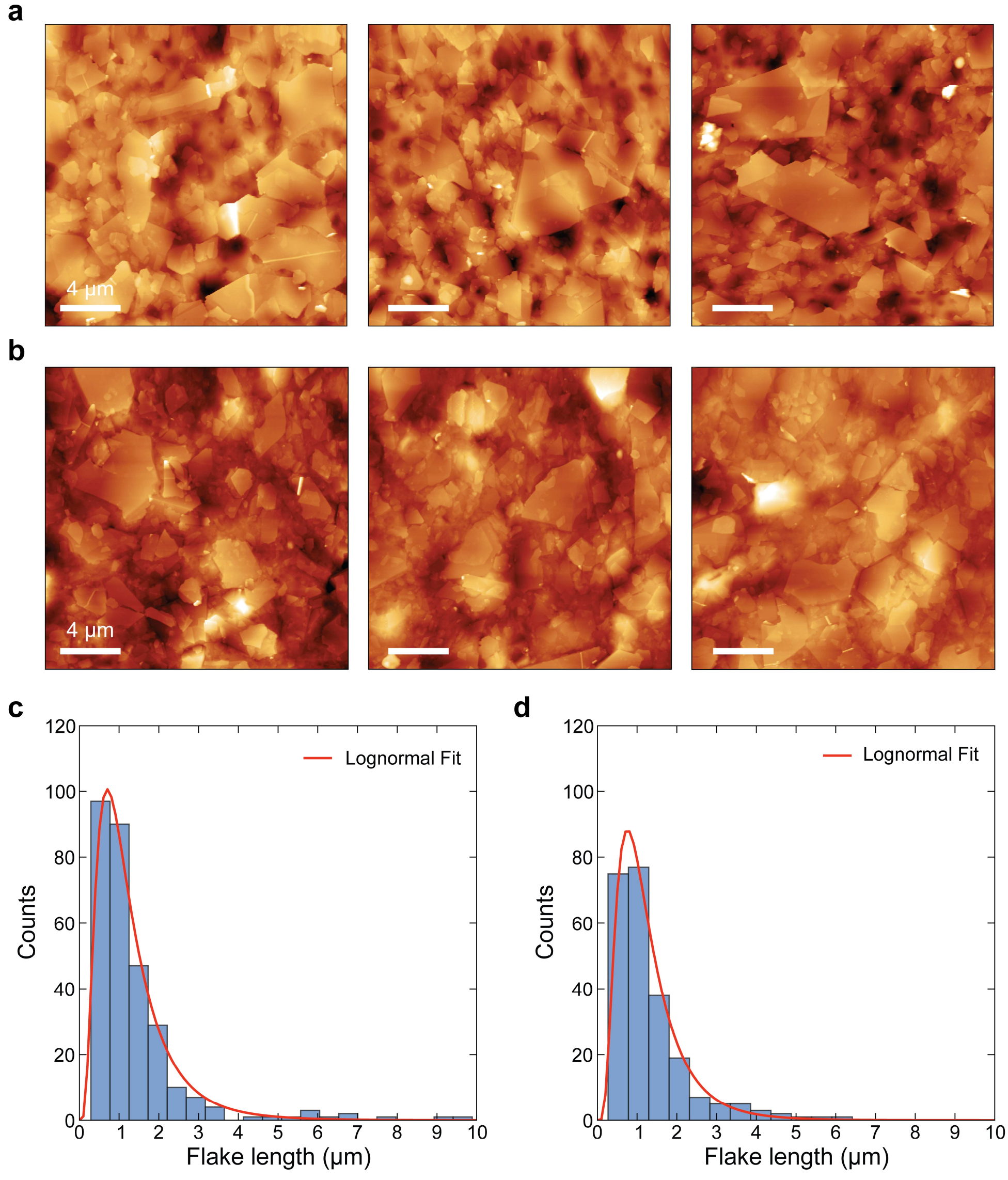

**Figure S5. AFM characterization of $MoS_2$ flakes and flake length distribution.** (a,b) Atomic force microscopy (AFM) topography images of $MoS_2$ flakes transferred onto (a) tattoo paper and (b) waterslide decal paper, acquired at three different locations on each substrate. (c,d) Corresponding flake length distributions obtained from the AFM images for (c) tattoo paper and (d) waterslide decal paper. Solid red lines represent lognormal fits used to extract the statistical parameters (mode and mean) of the flake lengths.

## Raman characterization of transferred $MoS_2$ flakes

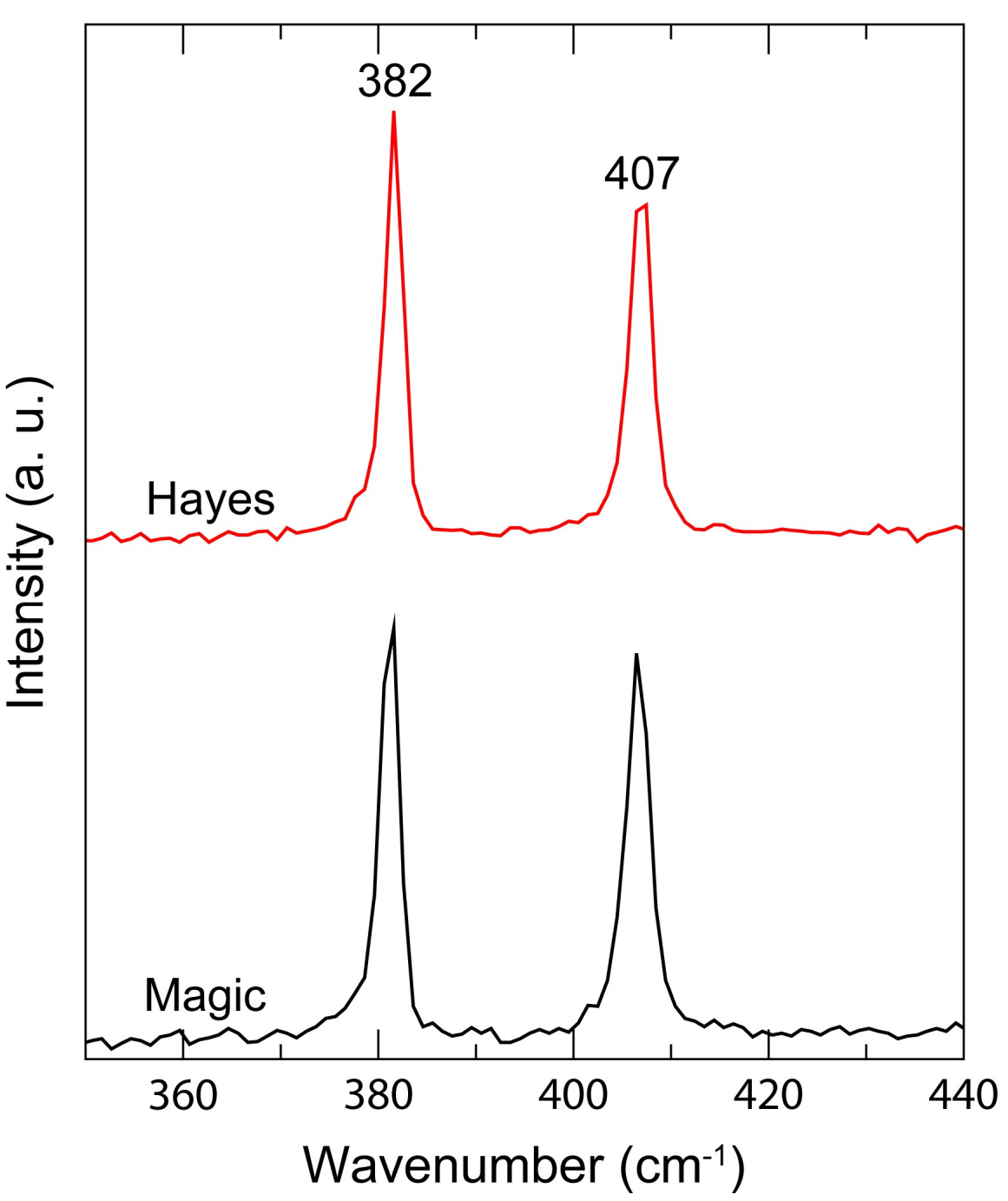

**Figure S6. Raman characterization of transferred $MoS_2$ nanosheets.** Raman spectra of $MoS_2$ nanosheets after transfer onto tattoo paper (Magic) and waterslide decal paper (Hayes).

## Transfer Length Method (TLM) analysis of $MoS_2$ films on tattoo paper

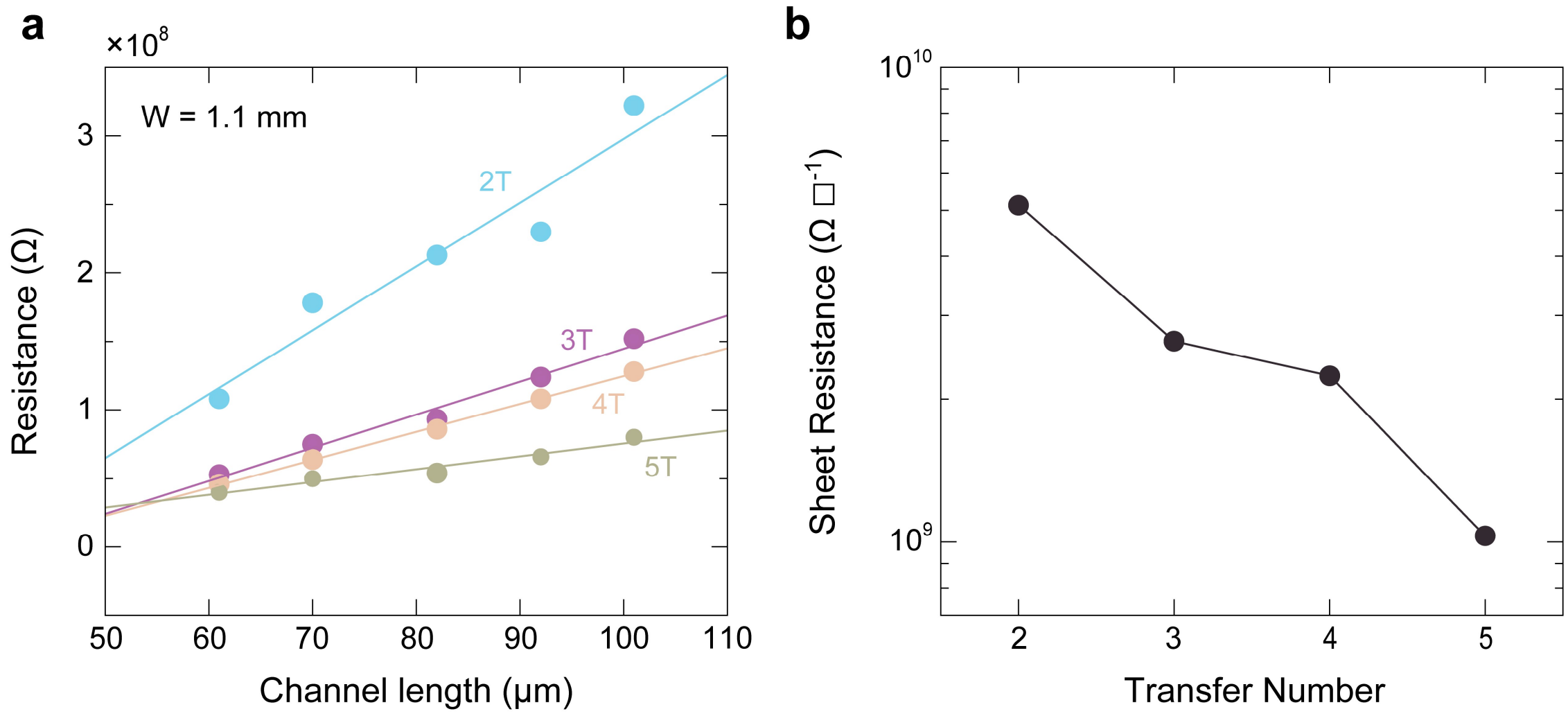

**Figure S7. TLM analysis and sheet resistance extraction for $MoS_2$ films on tattoo paper.** (a) TLM measurements for films consisting of different number of transfers. Solid lines correspond to the linear fits performed on each data set. (b) Extracted sheet resistance as a function of the number of transfers.

## Zoomed-in optical microscope image of $MoS_2$ tattoo device on curved metallic surface

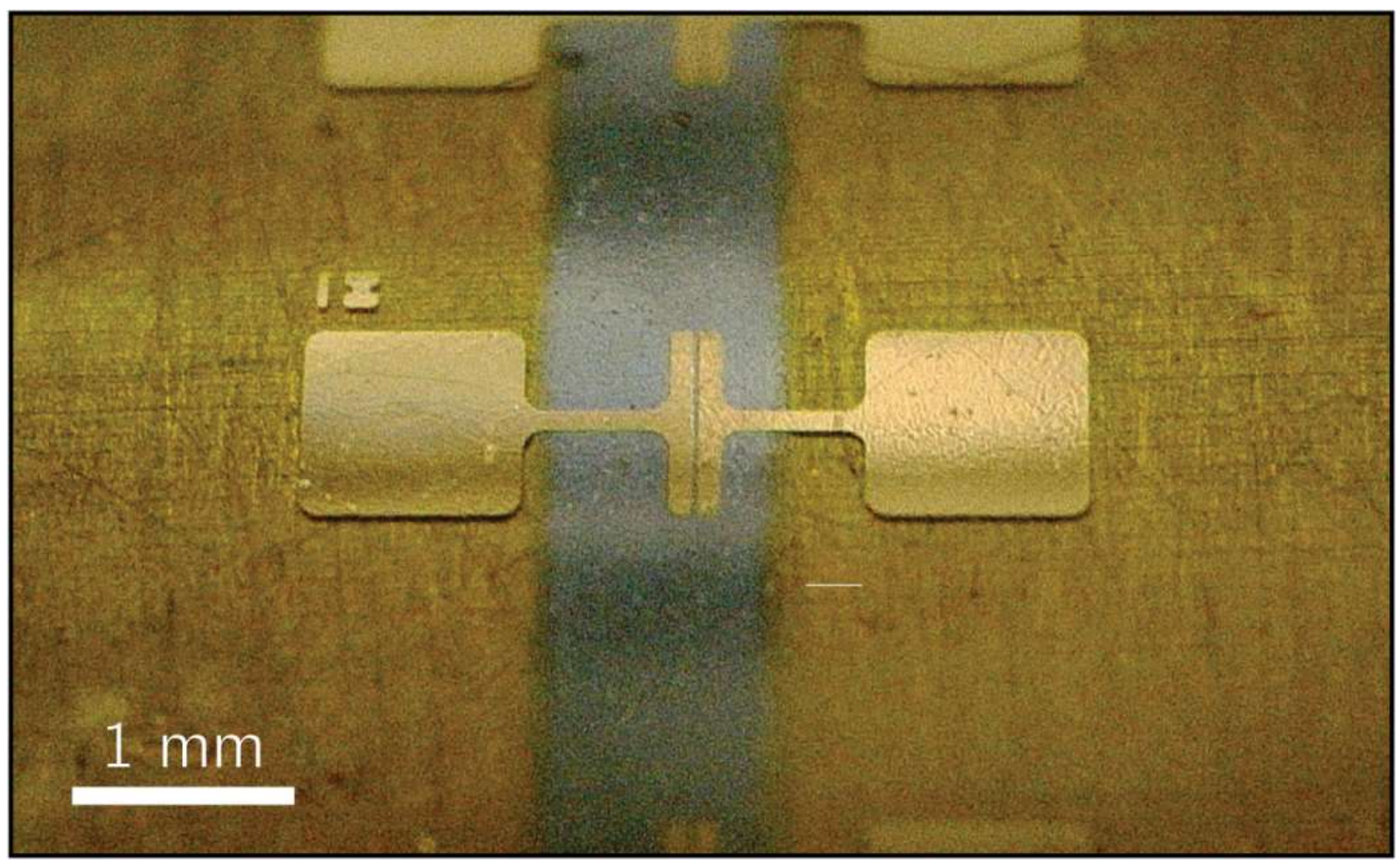

**Figure S8. Optical microscope image of $MoS_2$ device fabricated on tattoo paper after transfer onto a curved metallic surface.** The device shows high conformability to the curved surface.

## Scanning electron microscopy (SEM) image of a tattoo device on synthetic leather

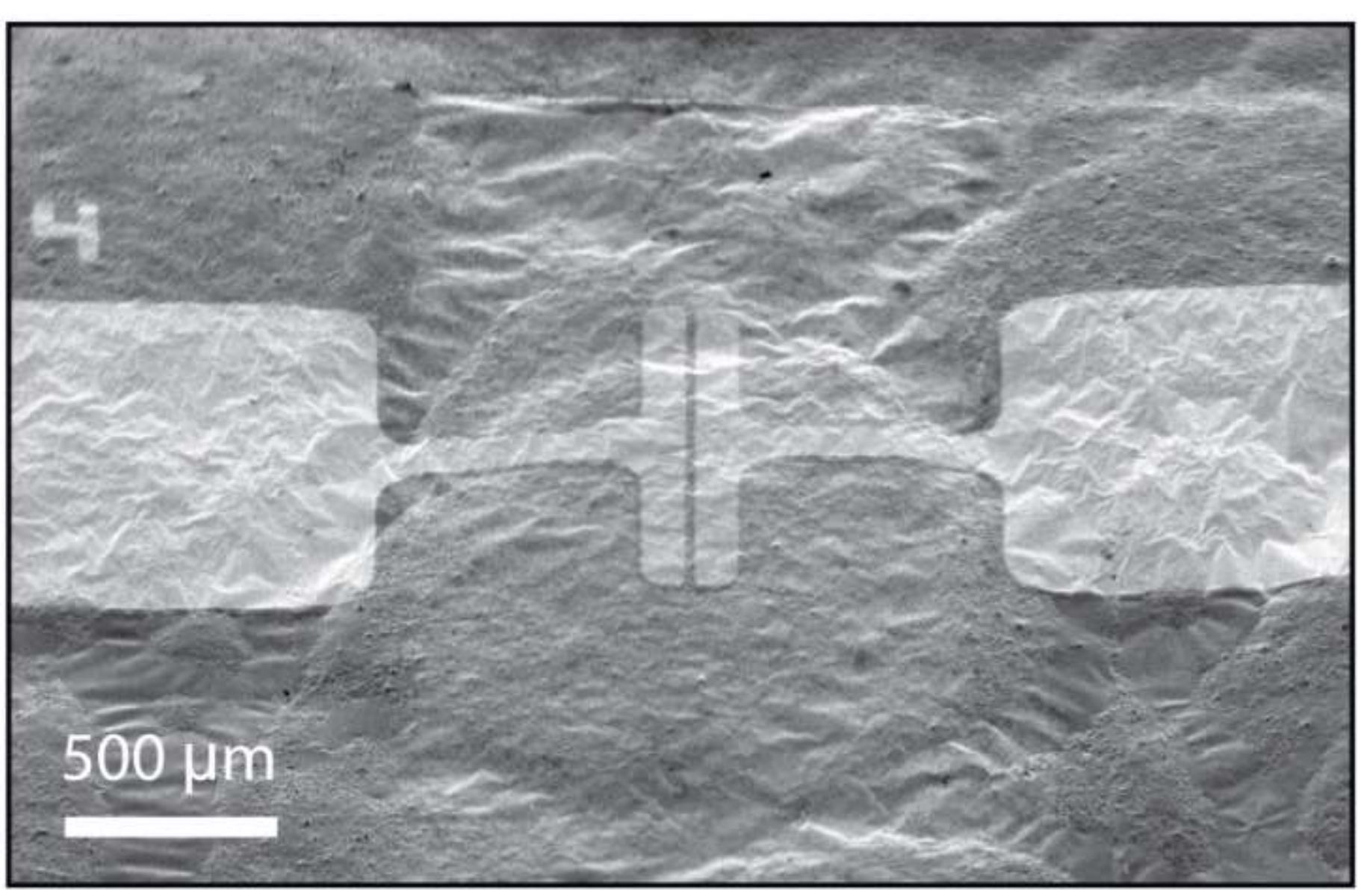

**Figure S9. Scanning electron microscopy (SEM) image of the conformable device after transfer onto synthetic leather.** The device conforms to the rough synthetic leather surface without compromising its structural integrity, demonstrating high conformal adhesion.

## Photoresponse characteristics of photodetectors on a curved metallic surface and synthetic leather

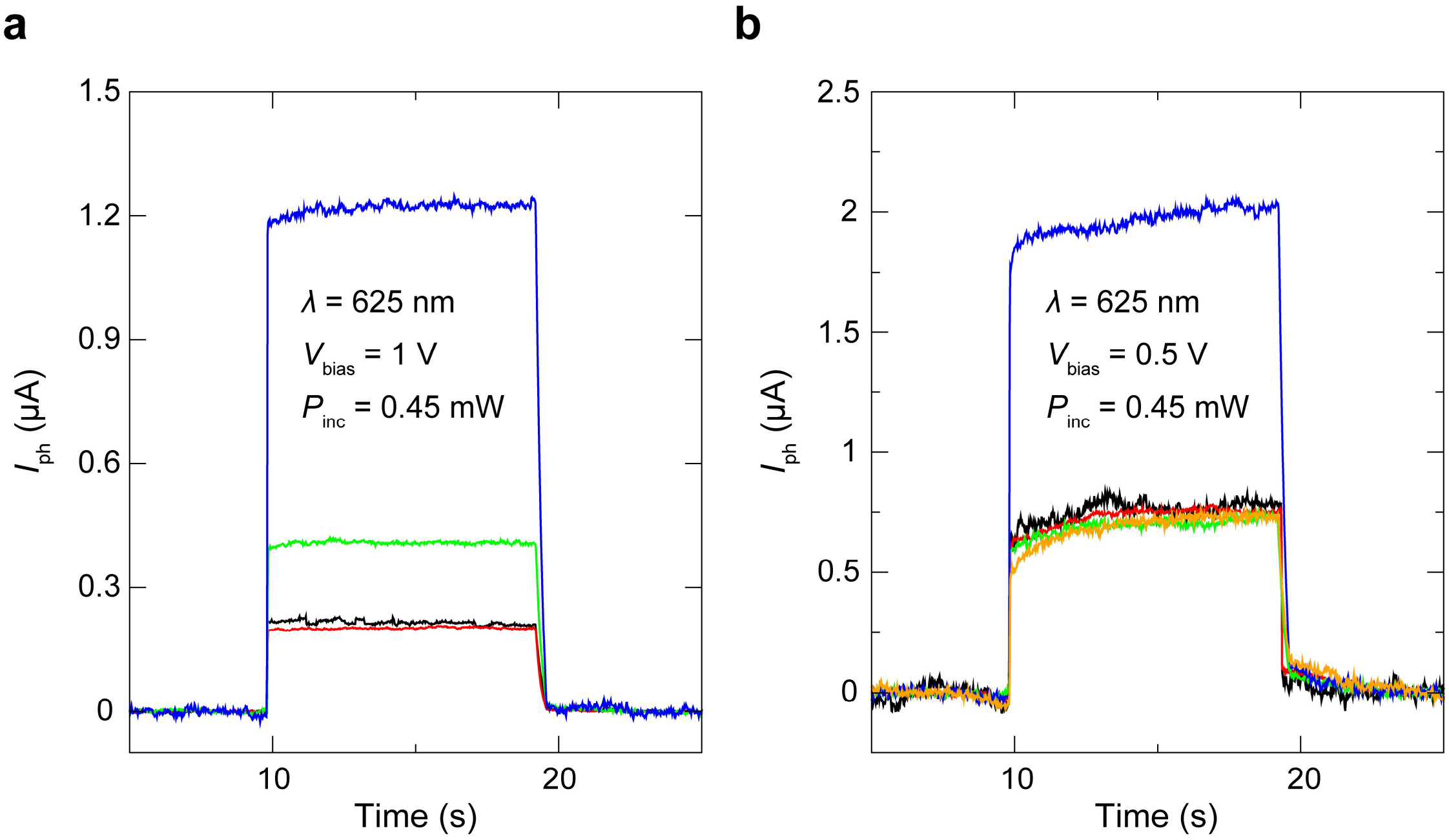

**Figure S10. Photoresponse of $MoS_2$ tattoo devices on a curved metallic and rough synthetic leather surface.** Time-resolved photocurrent measurements of $MoS_2$ photodetectors fabricated on waterslide decal paper after their transfer onto (a) a curved metallic surface and (b) synthetic leather.

## Electrical and photoresponse characteristics of a waterslide decal paper–based photodetector before and after transfer onto a leaf

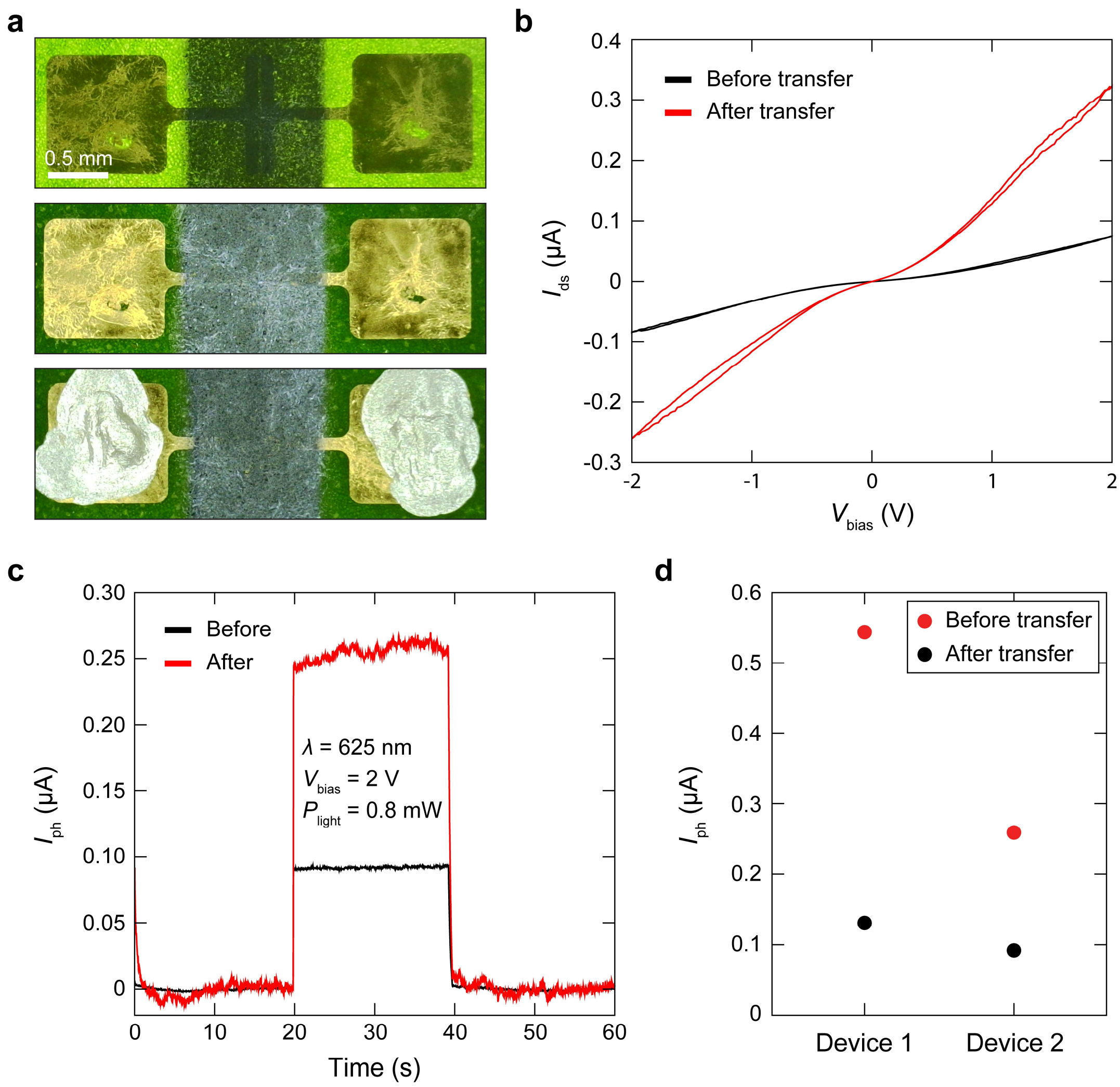

**Figure S11. Characterization of a photodetector fabricated on waterslide decal paper before and after transfer onto a leaf.** (a) Optical images of the device on a leaf shown (from top to bottom) in transmission mode, reflection mode, and after application of silver paste onto the gold pads to establish electrical contacts. (b) *I*–*V* characteristics and (c) time-resolved photocurrent response of a representative device, measured before transfer (on waterslide decal paper) and after transfer onto a leaf. (d) Photocurrent ($I_{ph}$) values obtained from two separate devices before and after transfer.

## Power-dependent photocurrent measurements for waterslide decal paper-based photodetectors on a leaf

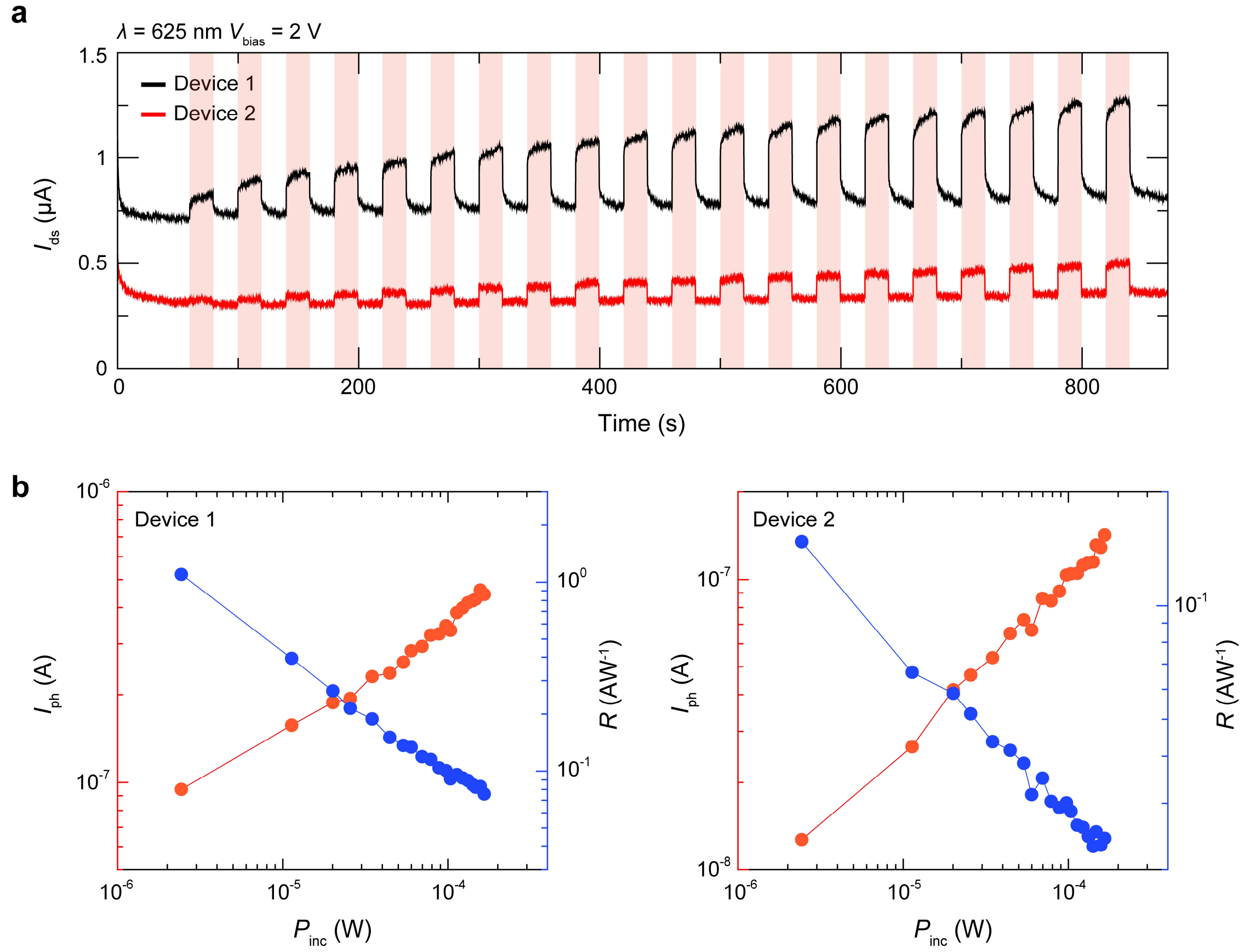

**Figure S12. Power-dependent photocurrent characterization for waterslide decal paper-based devices on a leaf.** (a) Time-resolved photocurrent measured during periodic ON/OFF illumination cycles under gradually increasing incident light power ($P_{inc}$). (b) Extracted $I_{ph}$ and $R$ as a function of $P_{inc}$ for the corresponding devices.

## Electrical and photoresponse characteristics of a tattoo paper–based photodetector before and after transfer onto a leaf

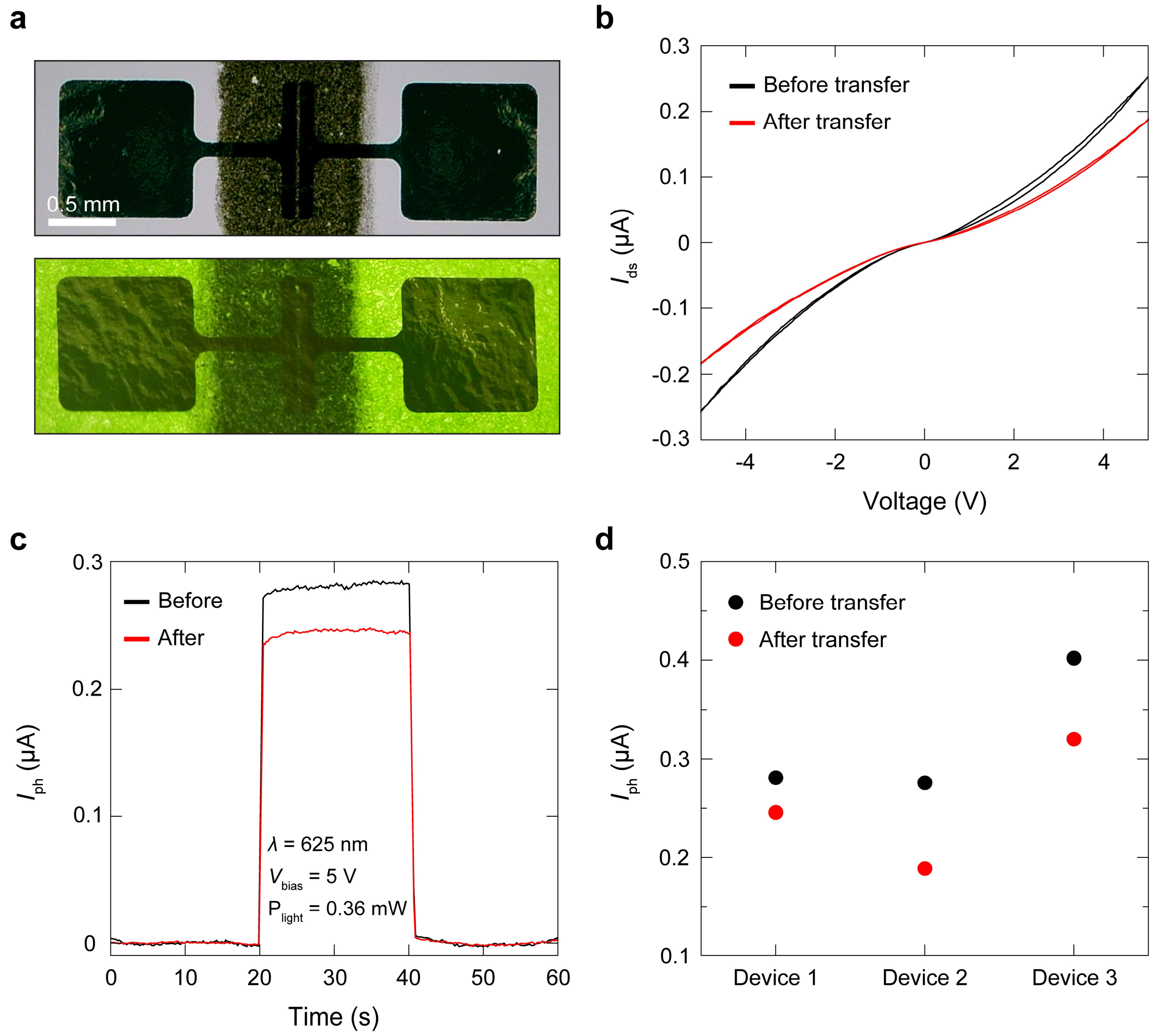

**Figure S13. Characterization of a photodetector fabricated on tattoo paper before and after transfer onto a leaf.** (a) Optical images of the device on tattoo paper (top) and after transfer onto a leaf (bottom). (b) *I*–*V* characteristics and (c) time-resolved photocurrent response of a representative device, acquired before transfer (on tattoo paper) and after transfer onto a leaf. (d) $I_{ph}$ values obtained from three separate devices before and after transfer.

## Power-dependent photocurrent measurements for tattoo paper-based photodetectors on a leaf

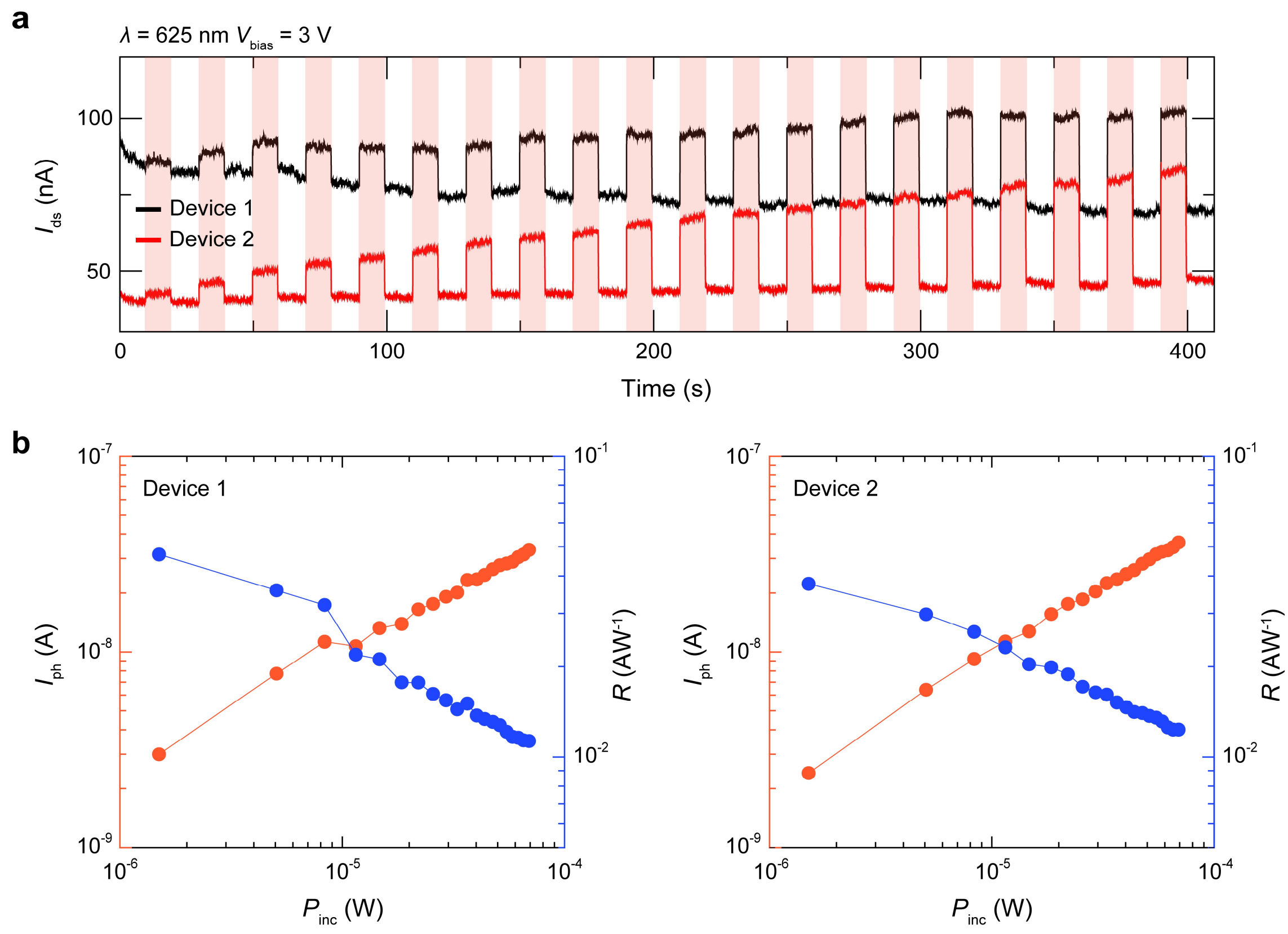

**Figure S14. Power-dependent photocurrent characterization for tattoo paper-based devices on a leaf.** (a) Time-resolved photocurrent measured during periodic ON/OFF illumination cycles under gradually increasing incident light power ($P_{inc}$). (b) Extracted $I_{ph}$ and $R$ as a function of $P_{inc}$ for the corresponding devices.

## Resistance as a function of temperature for devices after transfer onto synthetic leather

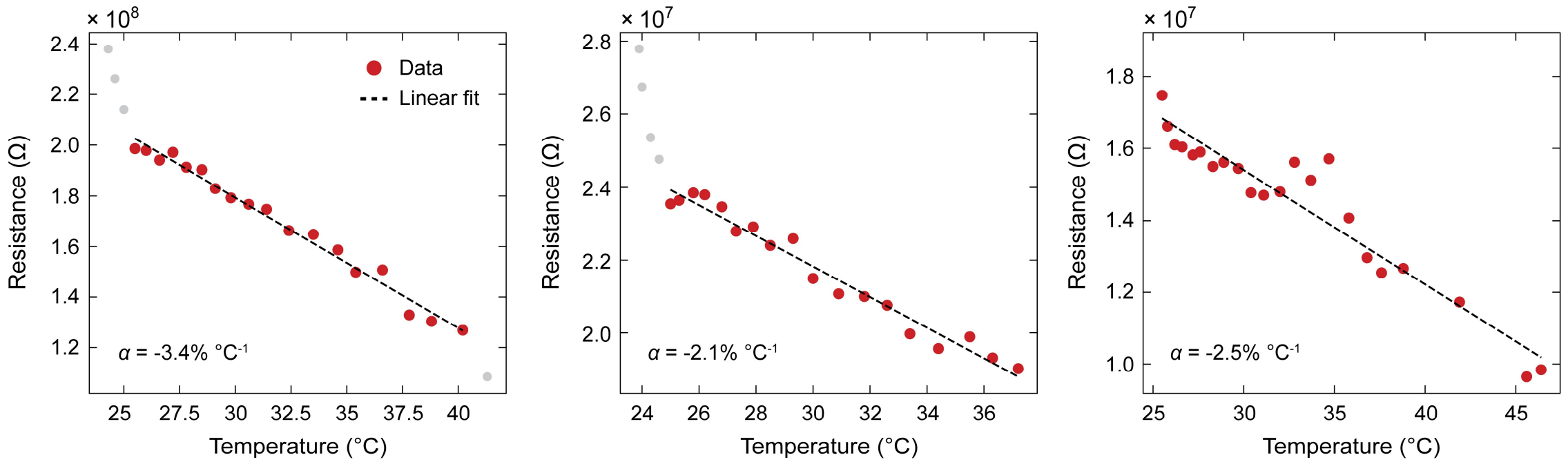

**Figure S15.** Electrical resistance as a function of temperature for $MoS_2$-based tattoo devices after transfer onto synthetic leather.

## Literature summary of reported temperature sensors

| **Sensing Material** | Temperature Range (°C) | TCR (% °C⁻¹) | Reference |
|---|---|---|---|
| XSBR/SSCNT-5 | 30 to 100 | −1.636 | 5 |
| Graphene-MPPU | 20 to 100 | −0.815 | 6 |
| NiO | 25 to 70 | −9.2 | 7 |
| PEDOT-TPU | 20 to 40 | 0.95 | 8 |
| graphene/PDMS | 25 to 75 | 0.8 | 9 |
| R-GO nanosheets in PU | 30 to 80 | 0.9 | 10 |
| Graphene | 25 to 85 | −1.48 | 11 |
| Graphene | 30 to 100 | −1.05 | 12 |
| $MoS_2$ | 20 to 60 | −1.94 | 13 |
| $MoS_2$ | 20 to 45 | 0.1 | 14 |
| $MoS_2$ | 27 to 85 | ~1 - 2 | 15 |
| rGO fiber | 30 to 80 | –0.21 ± 0.01 | 16 |
| Graphene nanowalls | 25 to 120 | 21.4 | 17 |
| $PtSe_2$ | 15 to 60 | −0.10 - −0.13 | 18 |
| $PtTe_2$ | 15 to 60 | 0.04 - 0.2 | 18 |
| Graphene | - | 0.1 - 0.3 | 19 |
| Aluminum | 0 to 60 | 0.314 | 20 |
| Polyaniline Nanofibers | 40 to 100 | 1.64 | 21 |
| Carbon black (CB) and Reduced Graphene Oxide | 20 to 60 | 0.6 | 22 |
| CNT-PEDOT:PSS | 22 to 45 | 0.68 | 23 |
| IGZO FN/SEBS | 35 to 75 | 2.1 | 24 |
| Nickel (Commercial) | - | 0.68 | 25 |
| Copper (Commercial) | - | 0.43 | 25 |
| Platinum (Commercial) | - | 0.39 | 25 |
| **$MoS_2$** | **24 to 41 ± 3** | **−2.1 - −3.5** | **This work** |

**Table S1: Summary of temperature sensors reported in the literature**. The table summarizes the temperature ranges over which the sensors were operated and the corresponding temperature coefficient of resistance (TCR) values. Apart from commercial sensors based on nickel, copper, and platinum, all other reported sensors have been fabricated on flexible or stretchable platforms.

## Key device performance metrics of ionic gel gated $MoS_2$ tattoo FETs on synthetic leather

| | $V_{ds}$ (V) | $V_g$ range (V) | μ ($cm^2 V^{-1} s^{-1}$) | $I_{off}$ (A) | $I_{on}$ (A) | $I_{ON/OFF}$ | $V_{th}$ (mV) | SS (meV $dec^{-1}$) |
|---|---|---|---|---|---|---|---|---|
| **Device 1** | 0.2 | −0.5 - 0.5 | 1.21 | $1.3\times10^{-8}$ | $3.8\times10^{-6}$ | $2.9\times10^{2}$ | 10 | 254 |
| **Device 2** | 0.2 | −0.5 - 0.5 | 0.23 | $5.5\times10^{-9}$ | $6.9\times10^{-7}$ | $1.3\times10^{2}$ | 28 | 315 |
| **Device 3** | 0.2 | −0.5 - 0.5 | 1.1 | $1.2\times10^{-8}$ | $3.5\times10^{-6}$ | $2.8\times10^{2}$ | -11 | 270 |
| **Device 4** | 0.2 | −0.5 - 0.5 | 2.43 | $3.9\times10^{-8}$ | $7.9\times10^{-6}$ | $2.0\times10^{2}$ | -20 | 296 |
| **Device 5** | 0.2 | −0.7 - 0.5 | 4.74 | $4.7\times10^{-8}$ | $1.6\times10^{-5}$ | $3.4\times10^{2}$ | -30 | 324 |
| **Device 6** | 0.02 | −0.7 - 0.5 | 17.6 | $2.2\times10^{-8}$ | $5.9\times10^{-6}$ | $2.7\times10^{2}$ | -41 | 314 |

**Table S2: Extracted device performance metrics for $MoS_2$-based tattoo FETs.** The table summarizes the key performance metrics, including mobility (μ), off-current ($I_{off}$), on-current ($I_{on}$), on/off ratio ($I_{on/off}$), threshold voltage ($V_{th}$), subthreshold slope (SS), along with the measurement parameters such as drain-source voltage ($V_{ds}$) and gate voltage range ($V_g$ range) used during the transfer curve measurements.

## Transfer characteristics of ionic gel gated $MoS_2$ tattoo FETs

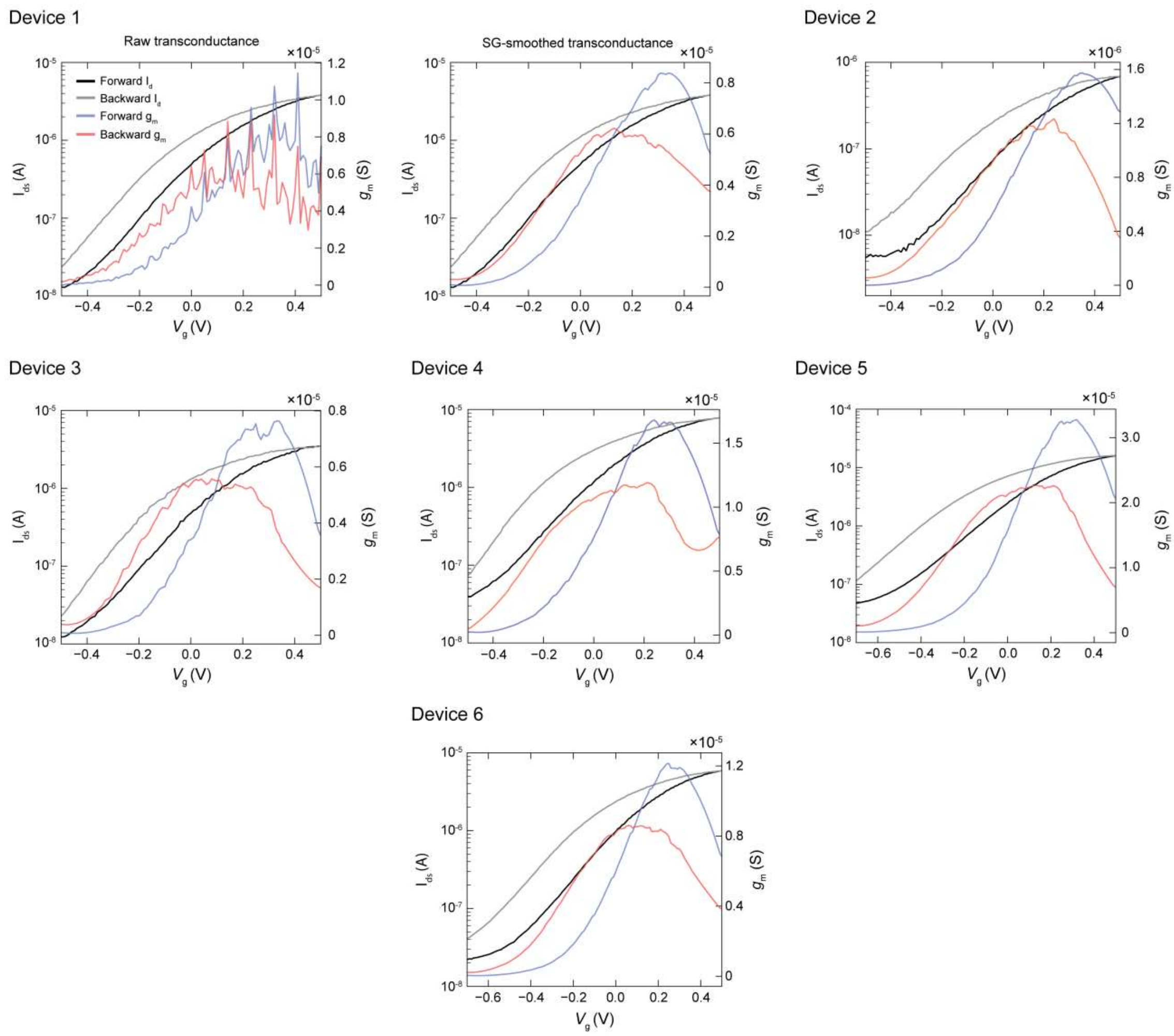

**Figure S16. Transfer characteristics of ionic-gel-gated $MoS_2$-based tattoo devices on synthetic leather.** The transfer curves, showing the evolution of the drain current ($I_{ds}$) during gate-voltage ($V_g$) sweeps, demonstrate consistent switching behavior across all devices, with only slight variations in performance metrics (see Table S2). Blue and red curves correspond to the extracted transconductance under forward and backward sweeps, respectively, for which a Savitzky–Golay filter was applied to smooth the data. For Device 1, both raw and smoothed transconductance data are shown to ensure transparent data processing.

## Linear-scale transfer characteristics of ionic gel–gated $MoS_2$ tattoo FETs

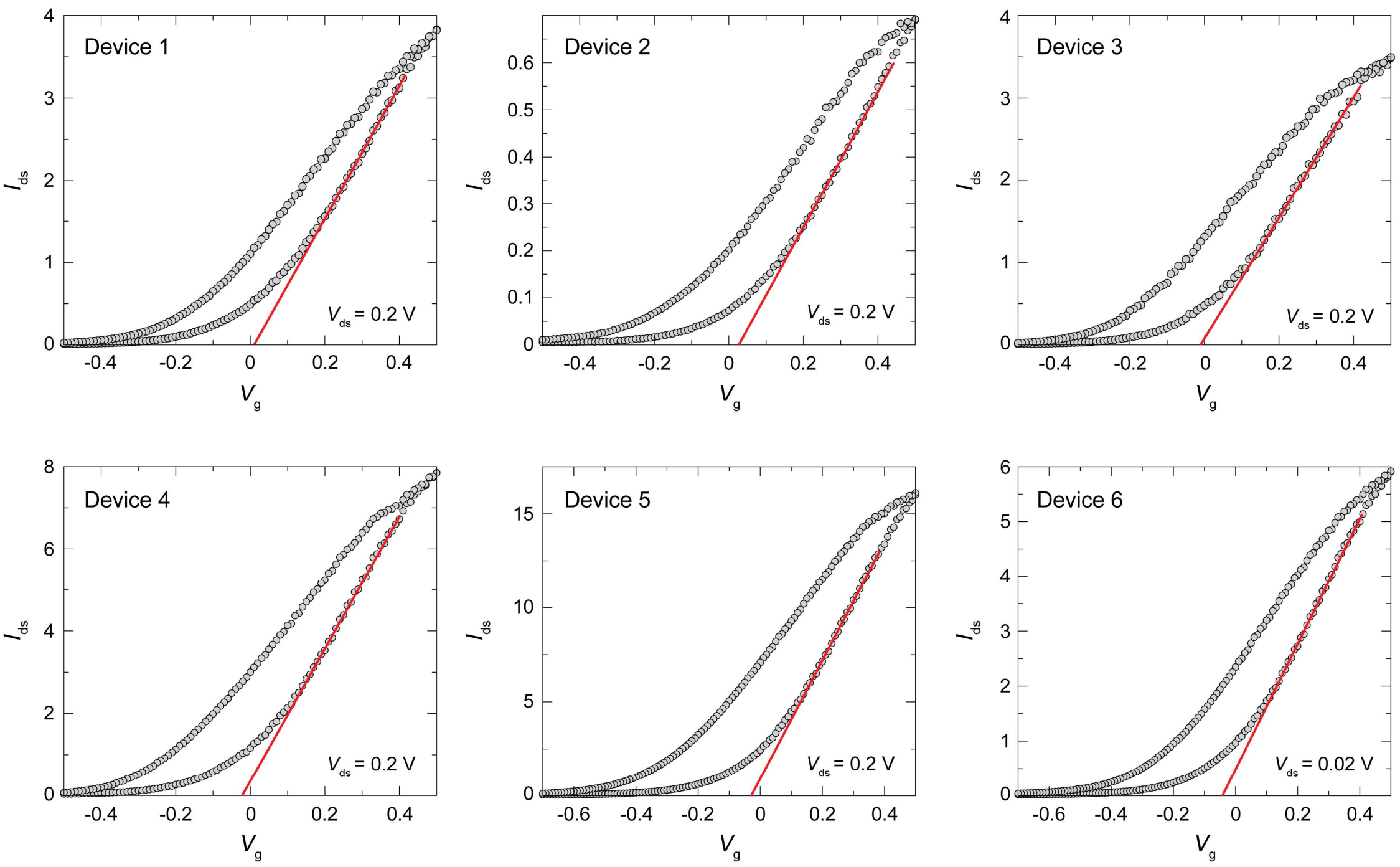

**Figure S17. Linear-scale transfer characteristics of ionic-gel-gated $MoS_2$-based tattoo devices on synthetic leather.** Linear-scale plots of the transfer curves presented in Figure S16. The red solid lines represent the linear extrapolation of the $I_{ds}$ - $V_g$ curve to the $V_g$-axis, used for the determination of $V_{th}$.

## Gating tattoo devices through ethylcellulose

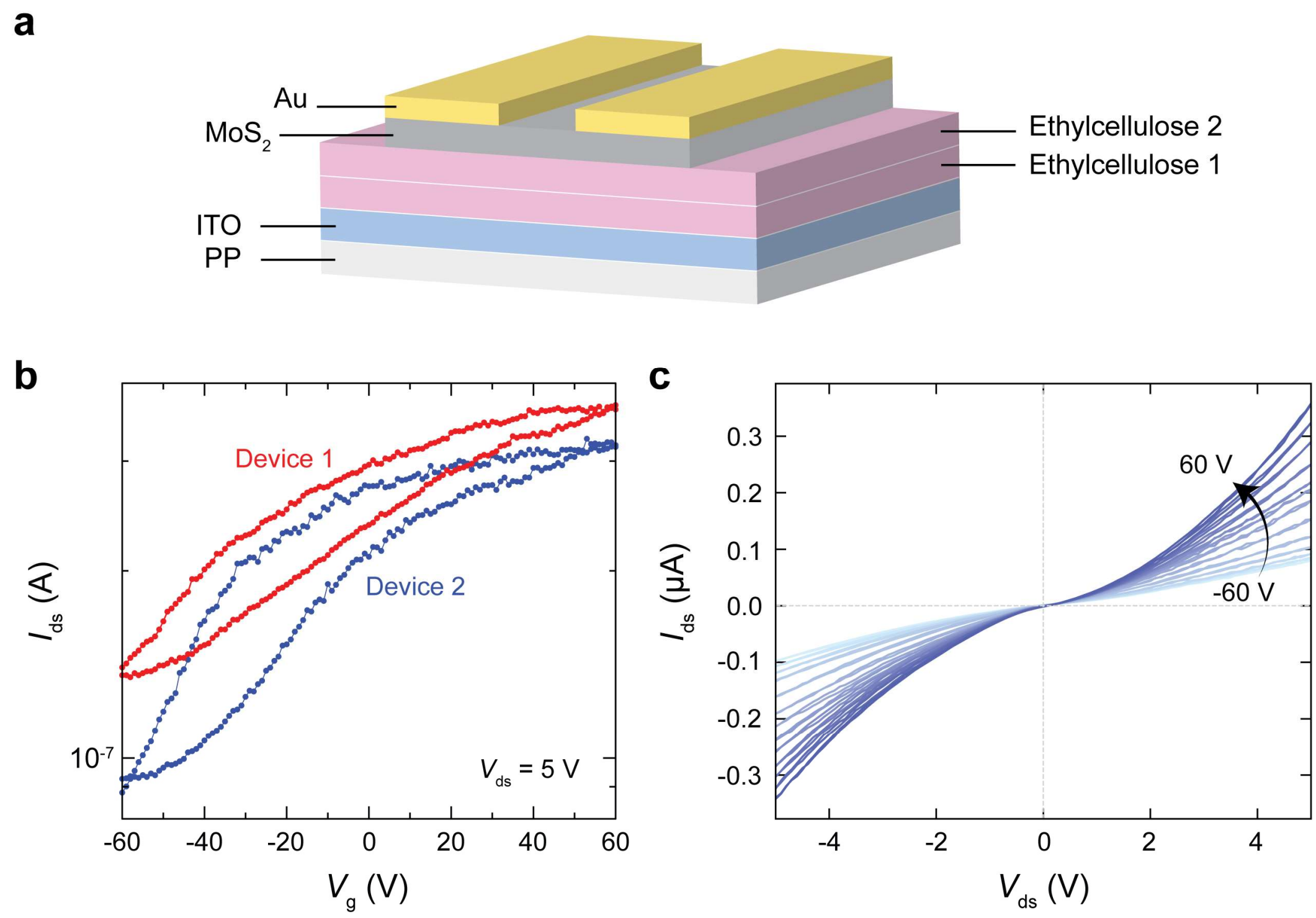

**Figure S18. FET characterization of $MoS_2$-based tattoo devices using ethylcellulose as the gate dielectric on an ITO-coated polypropylene (PP) substrate.** (a) Three-dimensional schematic of the device architecture, where two ethylcellulose layers serve as the gate dielectric between the $MoS_2$ channel and the ITO back-gate electrode. An additional bare ethylcellulose layer was first transferred onto the PP/ITO substrate prior to transferring the device to suppress leakage currents between the ITO gate and the source–drain contacts. Each transfer step was followed by annealing at 100 °C for 10 min. After completing the device structure, vacuum annealing was performed at 125 °C for 2 h. (b) Transfer characteristics measured from two representative devices, demonstrating gate-tunable channel conductivity. (c) Gate-dependent *I*–*V* characteristics of device 2 for gate voltages swept from −60 V to +60 V.

## Literature summary for FETs demonstrated on flexible and stretchable platforms

| Channel Material | Substrate | Gate dielectric | $V_g$ range (V) | μ ($cm^2 V^{-1} s^{-1}$) | $I_{on}/I_{off}$ | $V_{th}$ (V) | Ref. |
|---|---|---|---|---|---|---|---|
| $MoS_2$ | SU-8 | $Al_2O_3$ | -15 to 10 | ~16.2 ± 1.3 | ~$10^6$ | ~3.1 ± 0.4 | 26 |
| $MoS_2$ | PDMS | Ion gel | 0 to 1.8 | 0.4 – 1.4 | $10^4$ | ~1 | 27 |
| $MoS_2$ | Polyimide | $Al_2O_3$ | -10 to 14 | ~20 | $10^8$ | 1.1 | 15 |
| IGZO FN | Polyimide | Ion gel | -0.5 to 10 | 0.52 | ~$10^4$ | 3.7 | 24 |
| α-IGZO | PI/PEA/PUA | α-$Al_2O_3$ | -5 to 5 | 12.5 | > $10^7$ | 1.1 | 28 |
| $WSe_2$ | PET | Ionic liquid | -3 to 3 | 1.9 ± 0.4 | 2.9 × $10^3$ | 0.4 | 29 |
| $MoS_2$ | Leaf/Skin | Leaf/Skin | - | ~10 | $10^2$ | ~1* | 30 |
| Graphene | PET | Ion gel | -3 to 2 | 203 (h), 91 (e) | - | - | 31 |
| $ReS_2$ | PET | $Al_2O_3$ | -3.5 to 3.5 | 6.19 | ~$10^4$ | 0.44* | 32 |
| S-CNT | SEBS | NBR/SEBS | -8 to 3 | 20.2 | ~$10^4$ | ~ -3.1 | 33 |
| $MoS_2$ | Polyimide | Ion gel | 0 to 1.5 | 3.01 | ~$10^3$ | < 1 | 34 |
| $MoS_2$ | PET | $HfO_2$ | -8 to 8 | 13.9 ± 2 | > $10^5$ | - | 35 |
| IGZO:PTFE | Polyimide | $SiN_x/SiO_2$ | -30 to 30 | ~3.5 | ~$10^9$ | 3.95 | 36 |
| P3HT | Polyimide | Ion gel | -1.5 to 1.5 | 2.0 ± 0.7 | > $10^5$ | 0.5±0.1 | 37 |
| $WS_2$ | Polyimide | $Al_2O_3$ | -4 to 6 | 11 | > $10^6$ | - | 38 |
| $MoS_2$ | Parylene C – Polyimide | $Al_2O_3$ | -5 to 10 | 6.5 | ~$10^8$ | 3.8 ± 1.2 | 39 |
| 29-DPP-SVS | SEBS | SEBS-X-azide | -30 to 10 | 0.98 | $10^4$ | -1 | 40 |
| $MoS_2$ | Elastomer | $HfO_x$ | -2 to 5 | 2.1 | $10^5$ | 2.76 | 41 |
| $MoS_2$ | Polyimide | $Al_2O_3$ | -5 to 5 | 0.56 | $10^6$ | 0.125 | 42 |
| $MoS_2$ | Polyimide | PVF | -4 to 5.5 | 2.44 | $10^2$ – $10^3$ | 1.76 | 43 |
| $MoS_2$ | Polyimide | $Al_2O_3$ | -10 to 0 | 1.5 | 2.6×$10^4$ | −7.5 | 44 |
| 29-DPP-TVT and DPP-TTT | PVF | PVF | -6 to 2 | 0.098 | ~$10^2$ | 0.9 ± 0.2 | 45 |
| DNTT (p), PDI-8CN2 (n) | Parylene | Parylene | -5 to 2 (p), -5 to 5 (n) | 0.11 (p), 0.007 (n) | ~$10^5$ | −1.23 (p), −0.77 (n) | 46 |
| DNTT | Parylene | Parylene | -5 to 1 | 0.34 | ~$10^5$ | −1.72 | 47 |
| m-CNT–doped P3HT-NFs/PDMS | PDMS | Ion gel | -3 to 0 | 7.3 | 1.23×$10^4$ | -1.9 | 48 |
| **$MoS_2$** | **Ethylcellulose** | **Ion gel** | **-0.5 to 0.5** | **4.55 (avg) 17.6 (best)** | **2.5×$10^4$** | **0.011** | **This work** |

**Table S3: Literature summary of the figures of merit for FETs demonstrated on flexible and stretchable platforms.** The table gives detailed information regarding the channel material, substrates, dielectric layers, gate-voltage operation ranges, and key device parameters, such as mobility ($\mu$), on/off ratio ($I_{on}/I_{off}$), and threshold voltage ($V_{th}$). Some $V_{th}$ values marked with an "*" were derived by us via linear extrapolation of the linear region of the transfer curve, since they were not provided in the original paper.

**Supplementary Video Descriptions**

**Supplementary Video S1.** The video demonstrates the transfer process of devices fabricated on TheMagicTouch Tattoo 2.1 paper.

**Supplementary Video S2.** The video demonstrates the transfer process of devices fabricated on Hayes waterslide decal paper.

**Supplementary Video S3.** The video demonstrates the waterslide decal paper-based devices during stretching after transfer onto skin.

SUPPORTING INFORMATION REFERENCES